\documentclass[12pt]{article}
\usepackage{amssymb,amsmath,amsfonts}
\usepackage{graphicx}
\usepackage{pdfpages}
\graphicspath{{Figures/}}
\usepackage{color}

\usepackage{authblk}

\numberwithin{equation}{section}

\newcommand{\be}{\begin{equation}}
\newcommand{\ee}{\end{equation}}
\newcommand{\bea}{\begin{eqnarray}}
\newcommand{\eea}{\end{eqnarray}}

\newcommand{\av}[1]{\left\langle{#1}\right\rangle}

\newcounter{resultcounter}[section]

\newtheorem{thm}[resultcounter]{Theorem}
\newtheorem{lem}[resultcounter]{Lemma}
\newtheorem{prop}[resultcounter]{Proposition}
\newtheorem{cor}[resultcounter]{Corollary}
\newtheorem{definition}[resultcounter]{Definition}

\def\bed{\begin{definition}}
\def\eed{\end{definition}}

\renewcommand{\r}{{\rm R}}
\newcommand{\s}{{\rm S}}

\renewcommand{\i}{{\rm i}}

\def\qed{\hfill $\Box$\medskip}

\newcommand{\scalprod}[2]{\left\langle {#1}, {#2}\right\rangle}
\newcommand{\bbbone}{\mathchoice {\rm 1\mskip-4mu l} {\rm 1\mskip-4mu l}
{\rm 1\mskip-4.5mu l} {\rm 1\mskip-5mu l}}

\begin{document}

\title{Quantum Electron Transport in Degenerate Donor-Acceptor Systems}

\author[1]{Marco Merkli}
\author[2]{Gennady P. Berman}
\author[3]{Avadh Saxena\vspace*{.5cm}}

\affil[1]{\small Department of Mathematics and Statistics
	
	 Memorial University of Newfoundland
	 
	 St. John's, NL, Canada A1C 5S7
	 
	 merkli@mun.ca \vspace*{.5cm}}

\affil[2]{\small Theoretical Division, Los Alamos National Laboratory,
	
	and the New Mexico Consortium,
	
	 Los Alamos, NM 87545, USA
	 
	  gpb@lanl.gov \vspace*{.5cm}
  }

\affil[3]{\small Theoretical Division and Center for Nonlinear Studies,
	
	Los Alamos National Laboratory,
	
	Los Alamos, NM 87545, USA
	
	avadh@lanl.gov }

\maketitle

\hfill \small LA-UR-19-31360

\begin{abstract}
We develop a mathematically rigorous theory for the quantum transfer processes in degenerate donor-acceptor dimers in contact with a thermal environment. We calculate explicitly the transfer rates and the acceptor population efficiency. The latter depends critically on the initial donor state. We show that quantum coherence in the initial state enhances the transfer process. If the electron is initially shared coherently by the donor levels then the efficiency can reach values close to 100\%, while an incoherent initial donor state will significantly suppress the efficiency.  The results are useful for a better understanding of the quantum electron transport in many chemical, solid state, and biological systems with complex degenerate and quasi-degenerate energy landscapes.  	
\end{abstract}

\section{Introduction and main results} 

Electron transfer processes in chemical, physical and biological systems are often modeled based on the assumption of a {\em two-state donor-acceptor model}. However, degeneracies or near degeneracies in the energy of the donor and acceptor levels are brought about generically, for example, by space or spin coordinates \cite{Newton1, Newton2}, and generally by the complexity of the molecules exchanging the electron. This is the case, in particular, for electron transfer reactions in biomolecules  \cite{Blumberger,Wolynes} and chromophores in photosynthetic systems  \cite{Newton,JA}. To model this complexity, one should  replace the two-state model by a {\em two-level model} having degenerate energy levels.  The degeneracy of the donor and acceptor levels may be due to a complicated energy landscape with an effective potential exhibiting multiple minima at equal energies, but corresponding to different values of additional `coordinates'. In this situation, one may view those minima as different `sites' ({\em e.g.} spatial positions) where the electron can be localized, see Fig.\,1. One is then immediately lead to questions regarding the influence on the transfer process caused by quantum interference, coherence and localization or delocalization of the electrons (excitations) to be transferred. With regards to biology,  it was discussed in \cite{EG, TS, K} (and references therein) that degeneracy plays an
important role in the functional robustness and adaptability of biological systems. The notion of degeneracy is understood and used differently in various publications on that topic, but a common statement is that degeneracy leads to a significant decrease of fluctuations, rendering the performance of biological systems more stable.

\bigskip
\bigskip

\centerline{\includegraphics[width=11cm, angle=0]{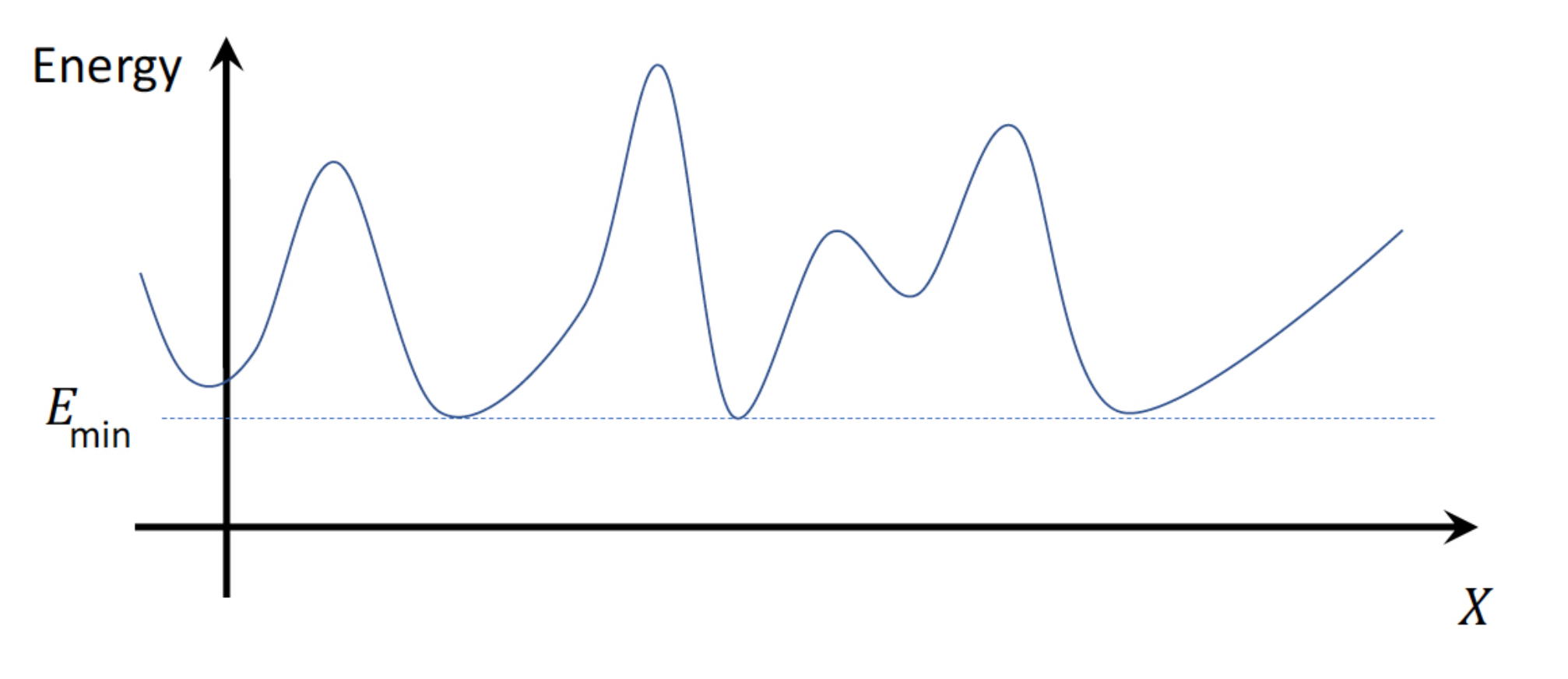}}

\medskip
\centerline{Fig.1:\ \ Energy landscape, three degenerate minima and generalized coordinate $X$.}
\bigskip

The goal of the current work is to analyze the dependence of the transfer process, such as its rate and efficiency (the amount of population transfer from $D \rightarrow A$), on the initial state and the number of degenerate states for $D$ and $A$. To start a systematic analysis of these questions, we propose to study here a {\em simple} mathematical model, in which the energies within the donor and the acceptor are exactly degenerate, the direct donor-acceptor matrix elements are chosen, for simplicity, to be the same between any donor and acceptor site, and where the DA complex is subject to the influence of a thermal environment. We use the formalism of quantum electrodynamics. In the current work, we consider a situation close to equilibrium, meaning that the DA complex is in contact with a single thermal reservoir. It is possible to extend our formalism to the non-equilibrium situation and include several reservoirs at different temperatures, giving rise to out of equilibrium stationary states with non-vanishing energy fluxes through the DA system connecting the reservoirs. It has been proposed in \cite{Cao} that such a setup can be used to exhibit experimentally the energy degeneracies in certain chemical compounds.

The parameters of the model are: the donor and acceptor energies $E_D$, $E_A$, with their respective degeneracies $N_D$ and $N_A$, the direct matrix element $V$ between any pair of donor and acceptor sites and the DA-environment interaction strength $\lambda$ and temperature, $T$.  Even though our  technique, the dynamical resonance theory, works as well for a DA system coupled strongly to the environment \cite{Dimer}, we consider in the current paper the parameter regime of {\em weak coupling to the environment}, characterized by
$$
\lambda^2 <\!\!< |E_D-E_A|,\quad \mbox{and}\quad \lambda^2 <\!\!< |V|.
$$  
Our main results are summarized as follows:
\begin{itemize}
\item[1.] We trace out the thermal environment and find the reduced density matrix of the DA system. We use a mathematically rigorous time-dependent perturbation theory in $\lambda$ and find the explicit form for each matrix element of the reduced DA density matrix, valid for all times $t\ge 0$, with an error $O(\lambda^2)$ which is independent of $t$ and $N_D$, $N_A$ (Theorem \ref{dynthm}). 

\item[2.]  We analyze the dynamics of the reduced DA density matrix and show that:

\begin{itemize}
\item[--] There is a manifold of explicit invariant states (Corollary \ref{cor1}).

\item[--] For large times, the DA density matrix approaches an explicit final stationary state which depends on the initial DA state, \eqref{m141}. 

\item[--] The dynamics of all DA reduced density matrix elements (populations and coherences) is irreversible, determined by explicitly calculated decay rates (Theorem \ref{dynthm} and Section \ref{explicitsection}). The decay rates are  independent of the initial DA state  and of $N_D$, $N_A$. Generically, the populations in the stationary state are not thermal (no Gibbs distribution) and coherences in the stationary state do not vanish.

\item[--]  Define the {\em transfer efficiency} to be the acceptor population in the stationary state (for large $t$), when starting out completely unpopulated. We show that the transfer efficiency depends critically on the quantum properties of the initial state. A coherent spread of the electron position over the donor sites in the initial state enhances the transfer efficiency dramatically, see Section \ref{secteff}.

\item[--]  The coupled DA reservoir dynamics leaves a two-dimensional DA space invariant. Namely, there are two fully symmetric states $|D\rangle$ and $|A\rangle$ (see \eqref{Dstate'}) such that if we take initial states of the form $\rho_{DA}\otimes\rho_{\r,\beta}$, where $\rho_{\r,\beta}$ is the reservoir equilibrium and $\rho_{DA}$ is any density matrix of the form $\rho_{DA} = a |D\rangle\langle D| +b |A\rangle\langle A| + c |D\rangle\langle A| + \overline c |A\rangle\langle D|$, then the following happens. The reduction to the DA system of the full state at time $t$ is again of the form $\rho_{DA}(t) = a(t) |D\rangle\langle D| +b(t) |A\rangle\langle A| + c(t) |D\rangle\langle A| + \overline c(t) |A\rangle\langle D|$.

However, if the initial DA state is not from this precise two-dimensional subspace, then the DA state $\rho_{DA}(t)$ explores all directions in Hilbert space and does not stay within the span of $\{|D\rangle, |A\rangle\}$. 

\item[--] We show that the fluctuation of the single donor site population (averaged over all sites) is proportional to $1/(N_D)^2$. This is in accordance with the central limit theorem and shows that bigger system size implies smaller fluctuations. This means that in our simple
degenerate donor-acceptor model, fluctuations in electron transfer are significantly suppressed for large systems. Even though we use a simple (and quantum mechanical) model, our findings coincide with those found in the literature on biological systems ({\rm e.g.} \cite{EG,TS,K}), as mentioned above in the introduction.

\end{itemize}

\item[3.] We outline in Section \ref{nosym} what happens in {\em quasi-degenerate} systems, where the donor and acceptor levels are not all at the same energies,  but may vary withing energy bands of size $\delta$ which are narrow compared to the size of the noise,  $\delta<\!\!<\lambda^2$. We argue that two time scales will emerge. On the first one, $\propto \lambda^{-2}$, the dynamics is very close to that corresponding to the degenerate situation. On a much larger second time scale, $\propto \lambda^2/\delta^2$, the DA system will feel the effect of the energy spread and converges to a final equilibrium state. This picture is supported by previous results, in \cite{MSB} where $N_D=1$ and $N_A=2$ was considered, and in \cite{GNB} where $N_D=1$, $N_A$ is general, but the noise is classical. A rigorous study of the quasi-degenerate regime is planned.
\end{itemize}

\bigskip

Let us now present the model and results in more detail. We consider $N_D$ donor states (sites) coupled to $N_A$ acceptor states (sites) via a direct matrix element $V$ and subject to the noise of a heat bath consisting of a collection of quantum harmonic oscillators, described by the total Hamiltonian 
\begin{equation}
	\label{hamiltonian}
H = H_\s +H_\r + \lambda G\otimes \varphi(h).
\end{equation}
The system Hamiltonian $H_\s$ and interaction operator $G$ are 
\begin{eqnarray}
H_\s &=& E_D\sum_{j=1}^{N_D}  |D_j\rangle\langle D_j| + E_A\sum_{k=1}^{N_A} |A_k \rangle\langle A_k| +V\sum_{j,k} \big( |A_k\rangle \langle D_j| + |D_j\rangle \langle A_k| \big),\label{m151.9}\\
G &=& g_D\sum_{j=1}^{N_D}  |D_j\rangle\langle D_j| + g_A\sum_{k=1}^{N_A}  |A_k \rangle\langle A_k|, 
\label{m152}
\end{eqnarray}
where $|D_j\rangle$ and $|A_k\rangle$ are the states in which the $j$th donor (site) and the $k$th acceptor (site) is populated, respectively and $E_D$, $E_A$, $V$ and $g_D$, $g_A\in\mathbb R$ are constants. The reservoir Hamiltonian is that of a field of independent harmonic oscillators, indexed for concreteness by $k\in{\mathbb R}^3$ (continuous modes), 
\begin{equation}
H_\r = \int_{{\mathbb R}^3} \omega(k) a^*(k)a(k) d^3k,
\label{resh}
\end{equation}
with dispersion $\omega(k) = |k|$.\footnote{It is not necessary in our approach to have a dispersion relation like this.  In the context of  molecular reservoirs, consisting of many protein or solvent atoms, the integral in \eqref{resh} is over a range of frequencies coupling to the DA system, weighted by a frequency density function. See for instance Appendix 5 of \cite{Dimer}.} The creation and annihilation operators satisfy the canonical commutation relations $[a(k),a^*(k')] = \delta (k-k')$.  The constant $\lambda$ in \eqref{hamiltonian} is the coupling parameter and $\varphi(h)$ is the field operator
\begin{equation}
\varphi(h) = \frac{1}{\sqrt  2}\int_{{\mathbb R}^3} \big(h(k) a^*(k) +\mbox{h.c.}\big) d^3k.
\label{m176}
\end{equation}
The form factor $h \in L^2({\mathbb R}^3, d^3k)$ is a square integrable function. The size of $h(k)$ determines how srongly the mode (oscillator) $k$ is coupled to the DA complex. Of course, \eqref{resh} and \eqref{m176} are the continuous versions of the discrete mode analogues
\begin{equation}
\label{danal}
H_\r= \sum_k \omega_k a^*_ka_k,\qquad \varphi(h) = \frac{1}{\sqrt  2}\sum_k \big(h_k a^*_k +\mbox{h.c.}\big),
\end{equation}
which are often used in the literature, and where the continuous mode limit is taken in quantities of interest after all. We start off directly with a continuous mode reservoir.

\medskip

The Hamiltonian $H_\s$, \eqref{m151.9}, describes a DA system with high symmetry, having the two properties:
\begin{itemize}
\item[(S1)] The energy of each site within the donor and the acceptor is constant, equal to $E_D$ and $E_A$, respectively. 

\item[(S2)] The direct matrix element between each donor and acceptor site is the same, $V$. 
\end{itemize}

This symmetry has direct consequences for the dynamics, which we explain in Sections \ref{symsect} and \ref{symcon}. We discuss in Section \ref{nosym} how the present situation can be viewed as a starting point for the analysis when the donor and acceptor energies fluctuate around the values $E_D$ and $E_A$ and so does the coupling $V$, and what to expect in this case.  
\bigskip
\bigskip

\centerline{\includegraphics[width=11cm, angle=0]{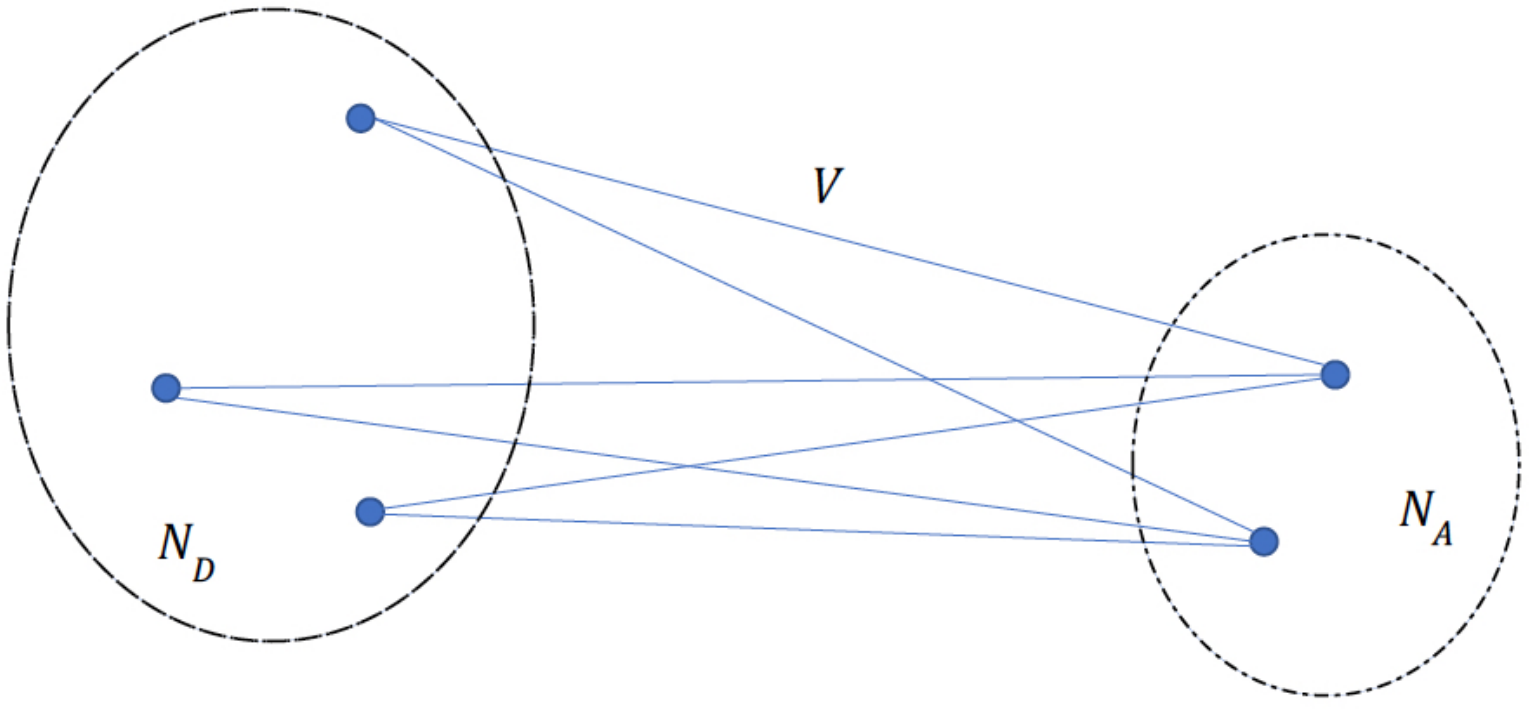}}

\medskip
\centerline{Fig.2:\ \ In $H_\s$, each donor site is coupled equally to each acceptor site.}




\subsection{Symmetry induced manifold of stationary states}
\label{symsect}

The Hamiltonian $H$, \eqref{hamiltonian} is block diagonal,
\begin{equation}
H =  H_{\rm eff}\oplus H_{D\perp}\oplus H_{A\perp}.
\label{m154}
\end{equation}
We now explain the three independent blocks. $H_{\rm eff}$ is the Hamiltonian of an effective dissipative two-level system with pure state space 
\begin{equation}
\label{hbars}
\bar{\mathcal H}_\s = {\rm span} \{ |D\rangle, |A\rangle\},
\end{equation}
spanned by the uniformly populated donor and acceptor states,
\begin{equation}
\label{Dstate'}
|D\rangle = \frac{1}{\sqrt{N_D}}
\sum_{j=1}^{N_D} |D_j\rangle , \qquad 
|A\rangle = \frac{1}{\sqrt{N_A}} \sum_{k=1}^{N_A} |A_k\rangle.
\end{equation}
The effective Hamiltonian is
\begin{eqnarray}
H_{\rm eff} &=& E_D |D\rangle\langle D| + E_A|A\rangle\langle A| + v \big(|A\rangle\langle D| + |D\rangle\langle A|\big) + H_\r\nonumber\\
&& +\lambda\big( g_D |D\rangle\langle D| + g_A  |A\rangle\langle A|\big)\otimes\varphi(h),
\label{m153} 
\end{eqnarray}
where the effective direct coupling matrix element is given by
\begin{equation}
v = V\sqrt{N_D N_A}.
\label{m160}
\end{equation} 
It follows from \eqref{m153} that $H_{\rm eff}$ leaves the Hilbert space $\bar{\mathcal H}_\s$, \eqref{hbars}, invariant. In particular, we have the following. The DA density matrix at any time will be a state on the two-dimensional space $\bar{\mathcal H}_\s$ if the initial DA matrix is. For instance, if the donor is initially homogeneously populated, in the state $|D\rangle\langle D|$, then the DA density matrix at all times is simply a $2\times2$ matrix on $\bar{\mathcal H}_\s$, a mixture of pure states involving only $|D\rangle$ and $|A\rangle$. However, as soon as the initial state does not lie within this effective two-state subspace, whose invariance is protected by symmetry, the evolution of the DA system explores all parts of the Hilbert space. We thus introduce the following.
\begin{itemize}
	\item[(a)] ${\mathcal H}_{D\perp}$ is the space of all linear combinations of $\{ |D_1\rangle ,\ldots,|D_{N_D}\rangle \}$ {\em which are orthogonal to $|D\rangle$}. 
	
	\item[(b)] ${\mathcal H}_{A\perp}$  is the space of all linear combinations of  $\{ |A_1\rangle,\ldots,|A_{N_A}\rangle \}$ {\em which are orthogonal to $|A\rangle$}.
\end{itemize} 
The Hamiltonians $H_{D\perp}$ and $H_{A\perp}$ in \eqref{m154} have the form 
\begin{equation}
H_{X\perp} = E_X\bbbone_\s +H_\r +\lambda g_X\bbbone_\s\otimes \varphi(h),\qquad 
X=D,A.
\label{m155}
\end{equation}
They act, respectively, on the Hilbert spaces  ${\mathcal H}_{D\perp}\otimes{\cal F}$ and ${\mathcal H}_{A\perp}\otimes{\cal F}$, where $\mathcal F$ is the Hilbert space of the environment. The {\em polaron transformation} for quantum oscillators is given by conjugation with a unitary displacement operator, $T = \exp\{\sum_k \alpha_k a^*_k - {\rm h.c.} \} = e^{\sqrt{2}\, \i \varphi(\alpha)}$ (see \eqref{danal}), where $\alpha_k\in\mathbb C$. It is defined equally well for continuous mode systems. It is well known that, choosing $\alpha_k = -\i \lambda g_X h_k / \omega_k$ and denoting the resulting displacement operator by $T_X$, the Hamiltonians $H_{X\perp}$ in \eqref{m155} are unitarily equivalent to the uncoupled but renormalized reservoir Hamiltonians,
\begin{equation}
T_X H_{X\perp} T^*_X = H_\r +\big( E_X -\tfrac 12 \lambda^2 g_X^2 \|h/\sqrt\omega\|^2\big)  \bbbone_\r,\qquad X=D,A. 
\label{m156}
\end{equation}
(We explain more details for instance in the proof of Lemma \ref{lemma3} below.) Note that modulo the additive constant, the right side of \eqref{m156} is simply $H_\r$, the Hamiltonian of the uncoupled reservoir alone. This means that there exist a multitude of invariant states. Namely,  let $\rho_{\r,\beta}$ be the reservoir equilibrium state. Then $e^{-\i t H_\r}\rho_{\r,\beta}\,  e^{\i t H_\r}= \rho_{\r,\beta}$ and \eqref{m156} implies  
\begin{equation} 
e^{-\i t H_{X\perp}} \big( |\psi\rangle\langle \psi|\otimes T_X^* \rho_{\r,\beta} T_X \big) e^{\i t H_{X\perp}} =  |\psi\rangle\langle \psi|\otimes T_X^* \rho_{\r,\beta}T_X ,\qquad \forall \psi\in {\cal H}_{X\perp}.
\label{m157}
\end{equation}
This leads to the following result.

\begin{cor}[Stationary states due to symmetry]
	\label{cor1}
For $X=D,A$, introduce the displacement operators $T_X = e^{-\sqrt 2\i\lambda g_X \varphi(\i h/\omega)}$. Then
\begin{itemize}
\item[1.] All density matrices of the form $\rho_\s\otimes (T_X^*\rho_{\r,\beta}T_X)$, where $\rho_\s$ is any mixture of pure states taken from ${\cal H}_{D\perp}$, are stationary states. All density matrices of the form $\rho_\s\otimes (T_X^*  \rho_{\r,\beta} T_X)$, where $\rho_\s$ is any mixture of pure states taken from ${\cal H}_{A\perp}$, are stationary states. 

\item[2.] Due to the existence of multiple stationary states, the asymptotic state, as $t\rightarrow\infty$, depends necessarily on the initial state.
\end{itemize}
\end{cor}

The invariant reservoir part is
$$
T_X^*\rho_{\r,\beta}T_X = e^{\sqrt 2\i\lambda g_X \varphi(\i h/\omega)} \rho_{\r,\beta}e^{-\sqrt 2\i\lambda g_X \varphi(\i h/\omega)},
$$
in which the `naked' state $\rho_{\r,\beta}$ is `dressed' with excitations due to the interaction with the DA system (arbitrarily many additional excitations are created in the reservoir, all in the single particle wave function $\propto h/\omega$).
\medskip

The first point in Corollary \ref{cor1} identifies an invariant manifold of dimension $\dim{\cal H}_{D\perp} + \dim{\cal H}_{A\perp}=N_D+N_A-2$. The dimension of the total system is, of course, infinite. The stationary states identified above are brought about by the symmetry of the Hamiltonian. We will see that there is exactly one more stationary state (for $\lambda\neq 0$), which is the equilibrium state of the whole, interacting DA-reservoir complex.  The second point of Corollary \ref{cor1} is an obvious general fact for dynamical systems with multiple stationary states.

\subsection{An a priori consequence for the dynamics}
\label{symcon}

Due to the decomposition \eqref{m154} the propagator is block diagonal as well,
\begin{equation}
e^{-\i t H} = e^{-\i t H_{\rm eff}}P_{\rm eff} + e^{-\i t H_{D\perp}}P_{D\perp} + e^{-\i t H_{A\perp}}P_{A\perp}. 
\label{m158}
\end{equation}
Here, $P_{\rm eff}=|D\rangle\langle D| + |A\rangle\langle A|$ and $P_{D\perp}$, $P_{A\perp}$ are the orthogonal projections onto ${\cal H}_{D\perp}$  and ${\cal H}_{A\perp}$, respectively, defined in (a), (b) before \eqref{m155} above. Consider initial states
  \begin{equation}
  \label{n1}
  \rho_{\s\r}(0) = \rho_0 \otimes \rho_{\r,\beta},
  \end{equation} 
where $\rho_0$ is an arbitrary density matrix of the donor plus acceptor and $\rho_{\r,\beta}$ is the reservoir thermal equilibrium at temperature $T=1/\beta>0$. Take now for $\rho_0$  a mixture of pure states from the linear span of $\{|D_1\rangle,\ldots,|D_{N_D}\rangle\}$, so that in particular, only donor sites are populated.  Let ${\cal O}$ be an observable of the donor alone, meaning that the matrix elements of $\mathcal O$ involving any $|A_k\rangle$ vanish. Then we have 
\begin{eqnarray}
\rho_\s &=&  \big(|D\rangle\langle D| +P_{D\perp}\big) \, \rho_\s\,  \big(|D\rangle\langle D| +P_{D\perp}\big),\nonumber\\
{\cal O} &=&  \big(|D\rangle\langle D| +P_{D\perp}\big) \, {\cal O}\, \big(|D\rangle\langle D|  +P_{D\perp}\big). 
\label{m161}
\end{eqnarray}
The average of $\cal O$ at time $t$ is given by
\begin{equation}
\av{{\cal O}}_t = {\rm tr} \Big( e^{-\i t H} ( \rho_\s\otimes\rho_{\r,\beta})\, e^{\i t H} ({\cal O}\otimes\bbbone_\r)  
\Big).
\label{m162}
\end{equation}
Due to \eqref{m161} the propagators $e^{\pm i tH}$ in \eqref{m162} are sandwiched between projections,
\begin{equation}
\big( |D\rangle\langle D| +P_{D\perp}\big) e^{\pm\i t H}\big( |D\rangle\langle D| +P_{D\perp}\big) = |D\rangle\langle D| e^{\pm\i t H_{\rm eff}} |D\rangle\langle D| +e^{\pm\i t H_{D\perp}}P_{D\perp}.
\label{m161.1}
\end{equation}
This leads to four terms in \eqref{m162} (two for each propagator). The Hamiltonian $H_{D\perp}$ does not depend on any quantity describing the acceptor. Also, $\langle D| e^{\pm i t H_{\rm eff}} |D\rangle$ depends on acceptor quantities only through the energy $E_A$ and the effective direct coupling $v$ (see \eqref{m160}). Hence so does the right side of \eqref{m161.1} and thus as well \eqref{m162}. 

Note that $\av{|D_k\rangle\langle D_\ell| }_t$ gives the populations ($k=\ell$) and coherences ($k\neq\ell$) between donor sites. We have just shown the following result.
\begin{cor}
Suppose that the initial DA state $\rho_\s$ is supported entirely on the donor, i.e., $\langle A_k| \rho_\s|A_\ell\rangle=0 = \langle D_k| \rho_\s|A_\ell\rangle$ for all $k,\ell$. Then 
\begin{itemize}
\item[1.] The populations of all donor sites and the coherences between any two donor sites, depend on the acceptor only via $E_A$ and $v$, but are independent of $N_A$, for all times.

\item[2.] The asymptotic state of the donor is independent of $N_A$, but it depends on the initial DA state.
\end{itemize}
\end{cor}

In view of point 2, we investigate in this paper, in particular, how the initial donor state influences the transfer efficiency of the process ({\em i.e.}, how much population weight is transferred from an initially populated donor to the initially empty acceptor).

\subsection{Main result: Time evolution of the DA system}

Define
\begin{equation}
\bar H_\s = E_D |D\rangle\langle D| + E_A|A\rangle\langle A| + v \big(|A\rangle\langle D| + |D\rangle\langle A|\big),
\label{m163}
\end{equation}
which is the DA part of $H_{\rm eff}$ for $\lambda=0$ (see \eqref{m153}). Its diagonalization is
\begin{equation}
\bar H_\s = e_1|\varphi_1\rangle\langle\varphi_1| +e_2|\varphi_2\rangle\langle\varphi_2|, 
\end{equation}
where
\begin{eqnarray}
	e_1 &=& \tfrac12 \Big\{  E_D+E_A +\sqrt{(E_D-E_A)^2 +4v^2}   \Big\}, \nonumber\\
	e_2 &=&\tfrac12 \Big\{  E_D+E_A -\sqrt{(E_D-E_A)^2 +4v^2} \Big\}\label{e12'}.
\end{eqnarray}
and\footnote{Note that $\lim_{v\rightarrow 0} |\varphi_1\rangle\langle\varphi_1| = |D\rangle\langle D|$ and  $\lim_{v\rightarrow 0} |\varphi_2\rangle\langle\varphi_2| = |A\rangle\langle A|$. }
\begin{equation}
\label{m22}
\varphi_{1,2} = \frac{1}{\sqrt{v^2+(e_{1,2}-E_D)^2}} \Big( v|D\rangle\langle D| + (e_{1,2}-E_D) |A\rangle\langle A| \Big).
\end{equation}

\medskip

We consider a fixed, but weak coupling between the DA and the noise, 
\begin{equation}
	\label{m20'}
	\lambda^2<\!\!< e_1-e_2 = \sqrt{(E_D-E_A)^2 +4v^2}.
\end{equation}
The DA density matrix at time $t$ is obtained by the reduction of the full DA-reservoir state, 
\begin{equation}
\rho_t = {\rm Tr}_{\r} \big(e^{- i tH} (\rho_0\otimes\rho_\r) e^{i t H}\big), 
\end{equation}
where the partial trace is taken over the reservoir degrees of freedom. 

\medskip
Our main result is Theorem \ref{dynthm} below, which gives the DA density matrix $\rho_t$ at all times $t\ge 0$. It gives explicitly {\em every density matrix element} (populations and coherences) of $\rho_t$  for arbitrary initial DA states. For convenience, let us recall here the definition of the following projections.
\begin{itemize}
\item[1.] $P_{D\perp}$ is the projection onto ${\cal H}_{D\perp}$, which is the space of all pure DA states $|\psi\rangle$ which are linear combinations of $\{|D_k\rangle\}_{k=1}^{N_D}$ {\em and which are orthogonal to $|D\rangle$}, $\scalprod{\psi}{D}=0$.
 \item[2.] $P_{A\perp}$ is the projection onto ${\cal H}_{A\perp}$, which is the space of all pure DA states $|\psi\rangle$ which are linear combinations of $\{|A_\ell\rangle\}_{\ell=1}^{N_A}$ {\em and which are orthogonal to $|A\rangle$}, $\scalprod{\psi}{A}=0$.
\item[3.] $\bar P_\s$ is the projection onto the effective two-level Hilbert space, $\bar{\cal H}_\s = {\rm span}\{ |D\rangle, |A\rangle\} ={\rm span}\{ |\varphi_1\rangle, |\varphi_2\rangle\} $.
\end{itemize}
We also introduce the effective two-level equilibrium Gibbs state as
\begin{equation}
\bar\rho_{\s,\beta}=  \frac{e^{-\beta\bar H_\s}}{{\rm tr}\, e^{-\beta \bar H_\s}},
\label{m142}
\end{equation}
a $2\times 2$ density  matrix acting on $\bar{\cal H}_\s$.

\begin{thm}[Dynamics of the DA density matrix]
\label{dynthm}
Let $\rho_0$ be an arbitrary initial DA density matrix and set
\begin{equation}
	P_{k\ell } = |\varphi_k\rangle\langle \varphi_\ell|,
	\label{m140}
	\end{equation}
	where $\varphi_1$, $\varphi_2$ are the eigenvectors \eqref{m22}. The reduced donor-acceptor density matrix at time $t\ge 0$ is given by
	\begin{eqnarray}
	\lefteqn{
		\rho_t =  {\rm Tr}( \rho_0\bar P_\s)\   \bar \rho_{\s,\beta}   +P_{D\perp} \rho_0  P_{D\perp} +P_{A\perp} \rho_0 P_{A\perp} + 2{\rm Re} \ e^{i t \varepsilon_4^{(3)}}  P_{A\perp} \rho_0 P_{D\perp}}	\label{dynamics}\\
	&& + \frac{ e^{it \varepsilon_1^{(2)}}}{e^{-\beta e_1} +e^{-\beta e_2}} \Big[ e^{-\beta e_2} P_{11}\rho_0 P_{11}  - e^{-\beta e_2} P_{21}\rho_0 P_{12}  - e^{-\beta e_1} P_{12}\rho_0 P_{21} +e^{-\beta e_1} P_{22}\rho_0 P_{22} \Big]\nonumber\\
	&& + 2 {\rm Re} \ e^{it \varepsilon_1^{(3)}}  P_{22}\rho_0  P_{11} + 2{\rm Re}\ \sum_{s=1,2} e^{it \varepsilon_2^{(s)}} P_{D\perp} \rho_0 P_{ss}+2 {\rm Re}\ \sum_{s=3,4} e^{it \varepsilon_2^{(s)}}  P_{A\perp} \rho_0 P_{(s-2)(s-2)} \nonumber\\
	&& +O(\lambda^2),\nonumber
	\end{eqnarray}
	where the error is uniform in $t$ and $N_D$, $N_A$. The resonance energies $\varepsilon_j^{(s)}$ are complex numbers, all having strictly positive imaginary parts, satisfying ${\rm Im} \varepsilon_j^{(s)}\propto\lambda^2$. Their explicit values are given in Section \ref{explicitsection}.
\end{thm}

{\em Remarks.\ } {\bf (1)} The resonance energies $\varepsilon_j^{(s)}$ depend on the following parameters: the DA effective energies $e_1, e_2$, the coupling parameters $\lambda, g_D, g_A$, the reservoir spectral density $J(\omega)$ (see \eqref{J}) and the temperature. They do not depend on $N_D$ nor on $N_A$, nor do they depend on the initial state $\rho_0$. It follows that the speed of the process is independent of $N_D, N_A, \rho_0$. 

{\bf (2)} The `topology' of the error is understood as follows. Denote the main term on the right side of \eqref{dynamics} by $\rho_t'$, so that $\rho_t=\rho_t'+O(\lambda^2)$. Then for any DA observable $X$, we have ${\rm Tr}(\rho_t X) = {\rm Tr}(\rho'_t X)+ R_t(X)$, where $|R_t(X)|\le C\lambda^2 \|X\|$, with a constant $C$ independent of $t,X,N_D,N_A$.

{\bf (3)} Another expression for $\rho_t$ is obtained by defining $x_j = e^{-\beta e_j}$, $j=1,2$, and using that 
\begin{equation}
{\rm Tr}(\rho_0 \bar P_\s)\ \bar\rho_{\s,\beta}= \frac{x_1}{x_1+x_2}\big( P_{11}\rho_0 P_{11} + P_{12}\rho_0 P_{21}) +\frac{x_2}{x_1+x_2}\big( P_{21}\rho_0 P_{12} + P_{22}\rho_0 P_{22}).
\nonumber
\end{equation}
The right side of \eqref{dynamics} can be rewritten, 
\begin{eqnarray}
\lefteqn{
	\rho_t =  \rho_0}\nonumber\\
&& -\frac{1-e^{i t\varepsilon_1^{(2)}}}{x_1+x_2}\Big[ x_1 \{P_{22}\rho_0 P_{22} - P_{12}
\rho_0 P_{21}\}	+x_2 \{P_{11}\rho_0 P_{11} - P_{21}
\rho_0 P_{12}\} \Big] \nonumber\\
&&- 2 {\rm Re} \ (1-e^{it \varepsilon_1^{(3)}})  P_{22}\rho_0  P_{11}- 2{\rm Re} \ (1-e^{i t \varepsilon_4^{(3)}})  P_{A\perp} \rho_0 P_{D\perp} \nonumber	\\
&&  -2{\rm Re}\ \sum_{s=1,2} (1-e^{it \varepsilon_2^{(s)}}) P_{D\perp} \rho_0 P_{ss} - 2 {\rm Re}\ \sum_{s=3,4} (1-e^{it \varepsilon_2^{(s)}})  P_{A\perp} \rho_0 P_{(s-2)(s-2)} \nonumber\\
&& +O(\lambda^2).\label{dynamics''}
\end{eqnarray}
The form \eqref{dynamics''} of $\rho_t$ shows immediately that the main term on the right side reduces to $\rho_0$ for $t=0$. 
\medskip

One readily sees that the main term on the right side of \eqref{dynamics} has unit trace. As ${\rm Im}\varepsilon_j^{(s)} >0$, the formula \eqref{dynamics} is directly exhibiting the {\bf asymptotic state}, 
\begin{equation}
\lim_{t\rightarrow\infty}\rho_t = {\rm Tr}( \rho_0\bar P_\s)\   \bar \rho_{\s,\beta}     +P_{D\perp} \rho_0  P_{D\perp} +P_{A\perp} \rho_0 P_{A\perp} +O(\lambda^2).
\label{m141}
\end{equation}

\begin{prop}[Donor populations and coherences]
	\label{corprop}
Set
	\begin{equation}
	{}[\rho_0]_{ss'} = \scalprod{\varphi_s}{\rho_0 \varphi_{s'}},\qquad s=1,2 \label{i4''}
	\end{equation}
	where $\varphi_{1,2}$ are given in \eqref{m22}. Define also
	\begin{equation} 
	\alpha = \frac{e_1-E_D}{v}. \label{i4'}
	\end{equation}
For all $k,\ell=1,\ldots,N_D$, we have 
\begin{eqnarray}
\lefteqn{
\scalprod{D_k}{\rho_t D_\ell}  = \scalprod{D_k}{\rho_0 D_\ell} -(1-e^{i t\varepsilon_1^{(2)}})	\frac{1}{N_D} \frac{1-\alpha^2}{1+\alpha^2}\  \frac{e^{-\beta e_2} [\rho_0]_{11} - e^{-\beta e_1} [\rho_0]_{22}}{e^{-\beta e_1}+e^{-\beta e_2}}} \label{m172} \\
&& -\frac{2}{N_D} \frac{|\alpha|}{1+\alpha^2}\,  {\rm Re} (1-e^{i t\varepsilon_1^{(3)}}) [\rho_0]_{21}\nonumber\\
&& -\sum_{s=1,2} \Big[ (1-e^{it \varepsilon_2^{(s)}}) \scalprod{D_k}{P_{D\perp} \rho_0 P_{ss}D_\ell} + (1-e^{-it (\varepsilon_2^{(s)})^*}) \scalprod{D_k}{P_{ss} \rho_0P_{D\perp} D_\ell} \Big]\nonumber\\
&& +O(\lambda^2).
\nonumber
\end{eqnarray}
\end{prop}

{\em Remarks.\ } {\bf (1)} The variation of each single matrix element $\scalprod{D_k}{\rho_t D_k}$ during the tranfer process is typically $O(1/N_D)$ since $\sum_{k=1}^{N_D} \scalprod{D_k}{\rho_t D_k} = {\rm tr} \rho_t \lesssim 1$. Therefore only a macroscopic group of donor sites can undergo a significant change (of the order one, not $O(1/N_D)$) during the transfer process. The same holds for acceptor sites.


{\bf (2)} It follows from \eqref{e12'} that $\frac{e_1-E_D}{v}\frac{e_2-E_D}{v} = -1$, a relation which allows us to replace easily $\frac{e_2-E_D}{v}$ by $-1/\alpha$ in our calculations.

\medskip
It follows from \eqref{m172} that 
\begin{eqnarray}
\lim_{t\rightarrow\infty}\scalprod{D_k}{\rho_t D_\ell}  &=& \scalprod{D_k}{\rho_0 D_\ell} -\frac{1}{N_D} \frac{1-\alpha^2}{1+\alpha^2}\  \frac{e^{-\beta e_2} [\rho_0]_{11} - e^{-\beta e_1} [\rho_0]_{22}}{e^{-\beta e_1}+e^{-\beta e_2}} \nonumber\\
&& -\frac{2}{N_D} \frac{|\alpha|}{1+\alpha^2}\,  {\rm Re} [\rho_0]_{21} -\frac{1}{\sqrt{N_D}}  \scalprod{D_k}{P_{D\perp} \rho_0 D}  -\frac{1}{\sqrt{N_D}} \scalprod{D}{\rho_0P_{D\perp} D_\ell} \nonumber\\
&& +O(\lambda^2).
\label{m173}
\end{eqnarray}
The last two contributions to the main term on the right side of \eqref{m173} are obtained from the sum over $s$ in \eqref{m172} by using that 
$\sum_{s=1,2} P_{ss} = |D\rangle\langle D| + |A\rangle\langle A|$, so $ \sum_{s=1,2}P_{ss} |D_j\rangle = \frac{1}{\sqrt{N_D}}|D\rangle$.

\subsection{Process rates}
\label{explicitsection}

According to Theorem \ref{dynthm}, the rates determining the decay of the dynamics are the imaginary parts of the resonance energies $\varepsilon_j^{(s)}$. We give here their explicit expressions, which depend on the matrix elements of the interaction and on the spectral density of noise at the frequency zero and the transition frequency $|e_1-e_2|$ of the effective two-level system. More precisely, set 
\begin{equation}
\bar G = g_D |D\rangle\langle D| +g_A |A\rangle\langle A|
\end{equation}
and denote its matrix elements in the basis $\{\varphi_1, \varphi_2\}$ by 
\begin{equation}
{}[\bar G]_{ij} \equiv \scalprod{\varphi_i}{\bar G\varphi_j} = \frac{g_Dv^2 +g_A(e_i-E_D)(e_j-E_D)}{\sqrt{\big(v^2+(e_i-E_D)^2\big) \big( v^2+(e_j-E_D)^2\big)} }.
\label{r2}
\end{equation}
Also, introduce the {\em spectral density of the reservoir} $J(\omega)$ by\footnote{The definition is $J(\omega) = \sqrt{\frac{\pi}{2}} \tanh\big(\beta\omega/2\big) \widehat  C(\omega)$, where $\widehat C(\omega) = \frac{1}{\sqrt{2\pi}} \int_{{\mathbb R}} e^{-\i\omega t} C(t)dt$ is the Fourier transform of the symmetrized reservoir correlation function $C(t) = \frac12 [ \av{e^{\i t H_\r} \varphi(h) e^{-\i t H_\r}\varphi(h)}_\beta
	+ \av{\varphi(h)e^{\i t H_\r} \varphi(h) e^{-\i t H_\r}}_\beta ]$. Here, $\av{\cdot}_\beta$ is the average in the reservoir thermal state.}
\begin{equation}
J(\omega) = \frac{\pi}{4}\omega^2 \int_{S^2} |h(\omega,\Sigma)|^2 d\Sigma, \qquad \omega\ge 0
\label{J}
\end{equation}
and define  
\begin{equation}
\label{Jtilde}
\widetilde J(0)=\lim_{\omega\rightarrow 0_+}\frac{J(\omega)}{\omega}. 
\end{equation}
The explicit expressions for the resonances appearing in \eqref{dynamics} are:
\begin{eqnarray}
\varepsilon_1^{(2)} &=& 4\i\lambda^2 \Big[ \frac{2}{\beta} \big( [\bar G]_{11}^2+[\bar G]_{22}^2 \big) \widetilde J(0) +[\bar G]^2_{12} \coth\big(\beta |e_1-e_2|/2\big) \ J(|e_1-e_2|)\Big], \nonumber\\
\varepsilon_1^{(3)} &=& e_1-e_2+\lambda^2( x_{12}+\i  y_{12}),\nonumber\\
\varepsilon_2^{(1)} &=& e_1-E_D + \lambda^2 x_1 +2\i\lambda^2\Big( \frac{1}{\beta}[\bar G]^2_{11}\widetilde J(0) +[\bar G]^2_{12}\frac{J(|e_1-e_2|)}{|1-e^{-\beta (e_1-e_2)}|} \Big), \qquad\nonumber\\
\varepsilon_2^{(2)} &=& e_2-E_D + \lambda^2 x_2 +2\i\lambda^2\Big( \frac{1}{\beta}[\bar G]^2_{22}\widetilde J(0) +[\bar G]^2_{12}\frac{J(|e_1-e_2|)}{|1-e^{\beta (e_1-e_2)}|} \Big), \nonumber\\
\varepsilon_4^{(3)} &=& E_D-E_A - \lambda^2 (E_D^2-E_A^2) \mu,
\label{res}
\end{eqnarray}
where we set
\begin{equation}
\mu = \frac{2}{\pi}\int_0^\infty \frac{J(\omega)}{\omega} \coth(\beta \omega/2) d\omega
\end{equation}
and where the real numbers $x_1$, $x_2$, $x_{12}$ and $y_{12}$ are given by
\begin{eqnarray}
x_1 &=& \big( g^2_D -[\bar G]_{11}^2)\mu,  \nonumber\\
x_2 &=&  \big( g^2_D- [\bar G]_{22}^2\big)\mu,  \nonumber\\
y_{12} &=& \frac{2}{\beta}  \big( [\bar G]^2_{11} + [\bar G]^2_{22} + 2[\bar G]_{12}^2\big)\widetilde J(0)  +2[\bar G]^2_{12}\coth\big(\beta|e_1-e_2|/2\big)\ J(|e_1-e_2|),\nonumber\\
x_{12} &=& \big( [\bar G]_{22}^2-[\bar G]_{11}^2 \big )\mu - [\bar G]_{12}^2\,  \frac{2}{\pi} \int_0^\infty \frac{J(\omega)}{\omega} e^{-\beta u}\coth(\beta\omega/2)d\omega \nonumber\\
&& -\frac{2}{\pi} [\bar G]_{12}^2 \ {\rm P.V.} \int_0^\infty J(\omega) \coth(\beta \omega/2) \Big( \frac{1}{u-e_1+e_2} -\frac{1}{u+e_1-e_2}   \Big) d\omega.
\label{m23}
\end{eqnarray}
The other two resonances appearing in \eqref{dynthm} are $\varepsilon_2^{(3)}$ and $\varepsilon_2^{(4)}$. They are obtained from the expressions of $\varepsilon_2^{(1)}$ and $ \varepsilon_2^{(2)}$ above in \eqref{res} by replacing $E_D$ with $E_A$ and $g_D$ with $g_A$.
\medskip

{\em Remark.\ } According to \eqref{res}, the relaxation rates ${\rm Im}\varepsilon_j^{(s)}$ only depend on the spectral density of noise $J(\omega)$ at the frequencies $\omega=0$ and $\omega=|e_1-e_2|$. Consequently, due to \eqref{J}, these rates only depend on the coupling function $h(\omega)$ at these two frequencies. {\em Nevertheless}, in order to be able to {\em derive} Theorem \ref{dynthm}  one must assume that $h$ is a square integrable function, {i.e.}, $\int_{{\mathbb R}^3} |h(\omega,\Sigma)|^2 \omega^2 d\omega d\Sigma<\infty$, for otherwise, the Hamiltonian \eqref{hamiltonian} cannot be  defined as an operator. In particular, the form factor $h$ must contain an {\em ultra-violet cutoff} to guarantee integrability for large values of $\omega$, even though this cutoff does not appear in the second order expressions (in $\lambda$) for the relaxation rates.

\subsection{Transfer efficiency}
\label{secteff}

The total donor population at time $t$ is given by
\begin{equation}
p_D(t) \equiv \sum_{k=1}^{N_D} \scalprod{D_k}{\rho_t D_k}.
\label{f5}
\end{equation}
Let $p_1,\ldots,p_{N_D}$ be a given probability distribution, $0\le p_j\le 1$, $\sum_{j}p_j=1$. We consider two families of initial states associated to  $\{p_j\}$:
\begin{itemize}
	\item[(inc)] The incoherent (classical) superposition $\rho_{\rm inc} =\sum_{j=1}^{N_D} p_j |D_j\rangle\langle D_j|$,
	
	\item[(coh)] The coherent (quantum) superposition  pure state $\rho_{\rm coh}=|\psi\rangle\langle\psi|$, where $|\psi\rangle = \sum_{j=1}^{N_D}\sqrt{p_j} |D_j\rangle$. 
\end{itemize}
For both choices (inc) and (coh), the initial donor population is $p_D(0)=1$. The incoherent $\rho_{\rm inc}$ may come about due to the prior contact of the donor with a decohering agent, making its density matrix diagonal. The coherent pure state $|\psi\rangle$ can be produced by applying to the donor molecule a short impulsive pulse of polarization $\hat e$, resulting in the initial donor state $|D_{\hat e}\rangle \propto {\hat e} \cdot \sum_j \vec\mu_j |D_j\rangle$, where $\vec\mu_j$ is the transition dipole moment vector of $D_j$ \cite{QUEBS}. Then $|D_{\hat e}\rangle\langle D_{\hat e}|$ is of the form $\rho_{\rm coh}$.
\medskip

The von Neumann entropy of the quantum state $\rho_{\rm inc}$ coincides with the {\em entropy of the probability distribution} $\{p_j\}$, given by $-\sum_jp_j\ln p_j$. The quantum state $\rho_{\rm inc}$, being pure, has zero von Neumann entropy. Nevertheless, we can view the entropy of $\{p_j\}$ as a measure for the coherence in $\rho_{\rm coh}$. It is maximal ($=\ln(N)$) for the uniform distribution $p_j=1/N_D$, $j=1,\ldots,N_D$, and it is minimal ($=0$) when exactly one $p_j$ is one and all others vanish. 
\medskip

\noindent
In this section we show the following.
\medskip

(1) The final donor population for the incoherent initial state is independent of how the donor is populated initially. For the initially coherent superposition the final donor population  depends on $\{p_j\}$ and is minimized (best transfer efficiency) for the uniform distribution, $p_j=1/N_D$, for all $j$, at which the entropy of the initial distribution $\{p_j\}$ is maximized.
\smallskip

(2) The final donor probability for the incoherent initial state is always {\em larger} or equal to that for the coherent initial state. Equality holds if and only if a single donor site is initially populated, i.e., for $\{p_j\}$ having minimal (= zero) entropy. We conclude that coherence in the initial state increases the final acceptor probability. 
\smallskip

(3) At large temperatures, $T>\!\!> e_1-e_2$, the final donor probability is at least $1-O(1/N_D)$ for large $N_D$ and so the transfer is suppressed. At low temperatures, $T<\!\! < e_1-e_2$, the final donor probability for the initially incoherent state is again at least $1-O(1/N_D)$ for large $N_D$. However, for the initially coherent state, it can reach values close to zero (perfect population transfer). 

\bigskip

To show these results, we start by using Proposition \ref{corprop} to get (modulo $O(\lambda^2)$),
\begin{eqnarray}
p_D(t) &=& p_D(0)-(1-e^{i t\varepsilon_1^{(2)}})	 \frac{1-\alpha^2}{1+\alpha^2}\  \frac{e^{-\beta e_2} [\rho_0]_{11} - e^{-\beta e_1} [\rho_0]_{22}}{e^{-\beta e_1}+e^{-\beta e_2}} \nonumber\\
&& -2 \frac{|\alpha|}{1+\alpha^2}\,  {\rm Re} (1-e^{i t\varepsilon_1^{(3)}}) [\rho_0]_{21}\nonumber\\
&& -2{\rm Re} \sum_{k=1}^{N_D} \sum_{s=1,2} (1-e^{it \varepsilon_2^{(s)}}) \scalprod{D_k}{P_{D\perp} \rho_0 P_{ss}D_k}. 
\label{m164}
\end{eqnarray}
This gives the asymptotic value
\begin{eqnarray}
p_D(\infty) &\equiv& \lim_{t\rightarrow\infty} p_D(t) \nonumber\\
&=&p_D(0)- \frac{1-\alpha^2}{1+\alpha^2} \frac{e^{-\beta e_2}[\rho_0]_{11} - e^{-\beta e_1} [\rho_0]_{22}}{e^{-\beta e_1}+e^{-\beta e_2}} - 2\frac{|\alpha|}{1+\alpha^2} \, {\rm Re} [\rho_0]_{12}\nonumber\\
&& - \frac{2}{\sqrt{N_D}} \sum_{k=1}^{N_D} {\rm Re} \scalprod{D_k}{P_{D\perp}\rho_0 D}.
\label{m165}
\end{eqnarray}

A direct calculation yields the following results.
\begin{itemize}
\item[(INC)] For the initial state $\rho_{\rm inc}$, \eqref{m165} becomes
\begin{eqnarray}
p_{D, \rm inc}(\infty)=1- \frac{1}{N_D} \frac{1}{1+\alpha^2} \frac{e^{-\beta e_1}\alpha^2 + e^{-\beta e_2}}{e^{-\beta e_1}+e^{-\beta e_2}}.
\label{m170}
\end{eqnarray}
The final donor population is {\em independent} of the distribution $\{p_j\}$. 
At high and low temperatures, \eqref{m170} reduces to
\begin{equation}
p_{D,\rm inc}(\infty)\approx 
\left\{  
\begin{array}{cl} \displaystyle  1- \frac{1}{2 N_D}
, & T>\!\!> e_1-e_2\\
\ &\ \\
\displaystyle 1 - \frac{1}{N_D (1+\alpha^2)}, & T <\!\!< e_1-e_2.
\end{array}
\right.
\label{m167}
\end{equation}

\item[(COH)] For the initial state $\rho_{\rm coh}$, we have 
\begin{equation}
p_{D,\rm coh}(\infty) =
 1- \frac{1}{N_D} \frac{1}{1+\alpha^2} \frac{e^{-\beta e_1}\alpha^2 + e^{-\beta e_2}}{e^{-\beta e_1}+e^{-\beta e_2}} \Big(\sum_{k=1}^{N_D} \sqrt{p_k}\Big)^2.
\label{m168}
\end{equation}
Now the final donor population depends on $\{p_j\}$. The Cauchy-Schwarz inequality gives 
\begin{equation}
\Big(\sum_{k=1}^{N_D} \sqrt{p_k}\Big)^2  \le \Big(\sum_{k=1}^{N_D} 1\Big) \Big(\sum_{k=1}^{N_D} p_k\Big) = N_D,
\label{CS}
\end{equation}
and equality holds in \eqref{CS} if and only if $p_k=1/N_D$ for all $k=1,\ldots,N_D$. This shows that the acceptor probability ($=1-p_{D, \rm coh}(\infty)$) is maximized for exactly one initial donor distribution, namely, the uniform one, in which the excitation is most delocalized. In particular, the transfer is most efficient for the distribution $\{p_j\}$ having {\em maximal entropy} ($=\log N_D$), namely the {\bf maximal final acceptor probability is given by}
\begin{equation}
 p_A^{\rm max} = 1-\min_{\{p_j\}} \ p_{D,\rm coh}(\infty) = \frac{1}{1+\alpha^2} \frac{e^{-\beta e_1}\alpha^2+  e^{-\beta e_2}}{e^{-\beta e_1}+e^{-\beta e_2}},
\label{m169}
\end{equation}
with the minimum over the $\{p_j\}$ is achieved uniquely for the uniform distribution. 

Since $\sqrt{p_k}\ge p_k$ we have
$(\sum_{k=1}^{N_D} \sqrt{p_k})^2\ge 1$ and equality holds if and only if exactly one of the $p_k$ equals one and all other ones vanish. Therefore, the relations \eqref{m170}, \eqref{m168} and \eqref{CS} show that 
$$
p_{D, \rm coh}(\infty) \le p_{D,\rm inc}(\infty)
$$
for any $\{p_j\}$, with equality if and only if $\{p_j\}$ is supported on a single site. This case is the one of {\em minimal (namely zero) entropy of $\{p_j\}$}. We conclude that coherence among the initial donor sites enhances the transfer efficiency.

At high and low temperatures, \eqref{m168} reduces to
\begin{equation}
p_{D,\rm coh}(\infty)\approx 
\left\{  
\begin{array}{cl} \displaystyle  1- \Big(\sum_{k=1}^{N_D} \sqrt{p_k}\Big)^2\frac{1}{2 N_D}
, & T>\!\!> e_1-e_2\\
\ &\ \\
\displaystyle 1 - \Big(\sum_{k=1}^{N_D} \sqrt{p_k}\Big)^2\frac{1}{N_D (1+\alpha^2)}, & T <\!\!< e_1-e_2.
\end{array}
\right.
\label{m171}
\end{equation}
Consider the low temperature regime with $p_k=1/\sqrt{N_D}$. Then $p_{D,\rm coh}(\infty)=1-\frac{1}{1+\alpha^2}$. For $\alpha=0$ we have total depletion of the donor, namely $p_{D, \rm coh}(\infty)=0$. What is the smallest value of $\alpha$? Setting
$$
\Delta:= E_D-E_A \ge 0,\quad \eta :=\frac{\Delta}{2v}\ge 0
$$
we get from \eqref{i4'} and \eqref{e12'} that
\begin{equation}
\alpha=\alpha(\eta) = -\eta +\sqrt{\eta^2 +1}.
\label{m175}
\end{equation}
The function $\eta\mapsto \alpha(\eta)$ is strictly decreasing and so it takes its minimum for $\eta\rightarrow\infty$, where $\alpha(\infty) =0$. The condition $T<\!\!< e_1-e_2 = \sqrt{\Delta^2+4v^2} = 2v\sqrt{\eta^2+1}$ becomes for large $\eta$ simply $T<\!\!<\Delta$. Therefore, in the regime  
$$
 0< v  <\!\!< \Delta,\quad T<\!\!< \Delta
$$
we have $p_{D,\rm coh}(\infty)\approx 0$. 
\end{itemize}
The dynamics of a donor coupled to acceptor levels in a related model are studied in \cite{GNB}.  There, a single donor level is coupled to $N_A$ acceptor levels at possibly different energy levels. The donor is coupled to each acceptor level by the same, scaled interaction ($V\rightarrow V/\sqrt N$ in \eqref{m151.9}, so $V$ is replaced in the Hamiltonian \eqref{m151.9} by $v$, see \eqref{m160}). The noise acts on each donor and acceptor level (is diagonal in the adiabatic DA basis), similar to \eqref{hamiltonian} and \eqref{m152}, however, in \cite{GNB}, the noise is {\em classical} (commutative), given by a stochastic process (telegraph noise). For this model and a degenerate acceptor ($N_A$ levels, all at the same energy, as in our situation), it is shown in \cite{GNB} that the final donor population is $1/2$. This coincides with our finding. Namely, for $N_D=1$ (as in that paper) and high temperature (which is believed to be modeled by classical noise), the donor population given in \eqref{m167} (or equivalently in \eqref{m171}) is also $1/2$.

\subsection{Population fluctuations }

In this section we use our results on the dynamics to show that the variance of the population of a single averaged  donor level is proportional to $1/(N_D)^2$, at each fixed time. This means that increasing the system (donor) size decreases the population fluctuations of the single donor level. We show how this stability property occurs in accordance with the central limit theorem. 
\medskip

In the previous sections, we have analyzed the averages (expectation values) of the donor population, for instance, \eqref{f5} is the average of the population of all donor sites and similarly,  ${\rm Tr}(\rho_t |D_k\rangle\langle D_k|)$ is that of site $k$ alone. The quantum measurement operator associated to the population of donor $k$ is the projection $|D_k\rangle\langle D_k|$. Let $X_k$ be the corresponding random variable, that is, the measurement outcome upon measuring the population of site $k$. Here $X_k$ takes the values $0$ or $1$ and its average and variance, at time $t$, are given by
\begin{eqnarray}
\langle X_k\rangle &=& {\rm Tr}(\rho_t |D_k\rangle\langle D_k| ) = \langle D_k ,\rho_t D_k\rangle, \label{f1.1}\\
{\rm Var}(X_k) &=&  \langle (X_k)^2\rangle  - \langle X_k\rangle^2 = \langle X_k\rangle  - \langle X_k\rangle^2.
\label{f1}
\end{eqnarray}
The last equality is due to $(X_k)^2=X_k$, as this random variable takes on the values $0$ and $1$ only, or equivalently, since $|D_k\rangle\langle D_k|$ is a projection. 
 The random variable 
\begin{equation}
\label{f13}
F_{N_D} = \frac{1}{N_D}\sum_{k=1}^{N_D} \big(X_k-\langle X_k\rangle\big)
\end{equation} 
is called the {\em fluctuation} of the single level population. It characterizes how much, averaged over all sites (as we take the weighted sum over $k$), the population of a single level deviates from its average value. $F_{N_D}$ has average zero and the {\em standard deviation}, which is the square root of its {\em variance}, measures by how much, typically, the site population deviates  from the average.

We now calculate the variance of $F_{N_D}$. Since its average vanishes, we have 
\begin{eqnarray}
\label{f14}
{\rm Var} \big( F_{N_D}\big) &=& \langle (F_{N_D})^2\rangle \nonumber\\\
&=& \frac{1}{(N_D)^2}{\rm Tr}\rho_t   \Big( \sum_{k=1}^{N_D} |D_k\rangle\langle D_k| -\langle X_k\rangle\Big)^2\nonumber\\
&=& \frac{p_D(t)\big(1-p_D(t)\big)}{(N_D)^2}.
\end{eqnarray}
The last equality holds since 
\begin{equation}
\label{f11.1}
\sum_{k=1}^{N_D} \langle X_k\rangle = p_D(t)
\end{equation} 
is the total donor population, defined in \eqref{f5}. Relation \eqref{f14} means that the population fluctuations decrease as a function of the size $N_D$ of the donor. Note that the relation \eqref{f14} holds in general -- we did not use the explicit form of $\rho_t$ or $p_D(t)$. Nevertheless, for the  specific model considered in this paper, we can use our explicit form for the donor dynamics, \eqref{m164}, and so we have an explicit expression for the variance of the fluctuation at each moment in time.

\medskip

{\bf Link to the central limit theorem.} For a sequence $Y_k$, $k=1,2,\ldots$ of independent random variables, having average $\langle Y_k\rangle$ and variance ${\rm Var}(Y_k)$, the central limit theorem states that \cite{Billing}
\begin{equation} 
\label{cl1}
\Big( \sum_{k=1}^{N} {\rm Var}(Y_k) \Big)^{-1/2} \sum_{k=1}^{N} \big( Y_k -\langle Y_k\rangle \big)\ \sim\  {\mathcal N}(0,1),
\end{equation}
where ${\mathcal N}(0,1)$ is a normal random variable with mean zero and variance one.\footnote{Note that in case ${\rm Var}(Y_k) =\sigma^2$ is the same for all $k$, \eqref{cl1} reduces to $\sum_{k=1}^{N} (Y_k-\langle Y_k\rangle) \sim \sqrt N \sigma  {\mathcal N}(0,1)$, or, $\frac 1N \sum_{k=1}^{N} (Y_k-\langle Y_k\rangle)\, \sim\, \frac{1}{\sqrt N}   {\mathcal N}(0,\sigma^2)$. This last form of the central limit theorem is maybe better known.} The $\sim$ means convergence in distribution, as $N\rightarrow\infty$. For \eqref{cl1} to hold the $Y_k$ are not only supposed to be independent but, strictly speaking, they must also satisfy a certain technical {\em Lyapunov condition}. 

Without checking this condition, nor addressing the independence of the $X_k$ introduced in \eqref{f1}, we can see what the result of the central limit theorem would imply when applied to $X_k$. Namely, \eqref{cl1} gives, for $Y_k$ replaced by $X_k$, 
\begin{equation}
\label{f12}
\sum_{k=1}^{N_D} \big( X_k -\langle X_k\rangle \big)\ \sim\  \Big( \sum_{k=1}^{N_D} {\rm Var}(X_k)\Big)^{1/2} \cdot \ {\mathcal N}(0,1).
\end{equation}
Next, from \eqref{f1},
\begin{equation}
{\rm Var}(X_k) =\langle X_k\rangle -\langle X_k\rangle^2 \le \langle X_k\rangle
\end{equation}
and so 
\begin{equation}
\sum_{k=1}^{N_D}{\rm Var}(X_k) \le \sum_{k=1}^{N_D} \langle X_k\rangle = p_D(t).
\label{f6}
\end{equation}
Dividing \eqref{f12} by $N_D$ on both sides, taking the variance on both sides and using the bound \eqref{f6} gives, for large $N_D$,
\begin{equation}
\label{f11}
\ {\rm Var}\big( F_{N_D}\big) = \frac{1}{(N_D)^2} \sum_{k=1}^{N_D}{\rm Var}(X_k)\le  \frac{p_D(t)}{(N_D)^2}.
\end{equation}
This finding, based on an application of the central limit theorem, is consistent with our exact formula \eqref{f14}. It shows in particular the correct scaling in $N_D$.

\medskip

{\bf Illustration.\ } Let us take the initial DA state to be $\rho_0=|D\rangle\langle D|$, in which all $N_D$ donor levels are populated, each with equal probability $1/N_D$, see \eqref{Dstate'}. Using that $\langle X_k\rangle=
\scalprod{D_k}{\rho_t D_k}$ and \eqref{m172}, we obtain\footnote{Note that the remainder term in \eqref{f6} is $O(\lambda^2/N_D)$, not just $O(\lambda^2)$ as one might infer from \eqref{m172}. This is so for the following reason. If instead of taking the observable $|D_k\rangle\langle D_k|$ in the trace in \eqref{f1}, we take $\sum_{k=1}^{N_D} |D_k\rangle\langle D_k|$, then our estimate for the size of the remainder is $O(\lambda^2)$, independent of $N_D$. This is due to the fact that the remainder estimate depends only on the norm of the observable $\sum_{k=1}^{N_D} |D_k\rangle\langle D_k|$ which equals one for all $N_D$. Hence the remainder for $\langle X_k\rangle$ in \eqref{f6} is $O(\lambda^2/N_D)$, since when summing it up $N_D$ times (for $k=1,\ldots,N_D$) we obtain a term $O(\lambda^2)$ independent of $N_D$. }
\begin{equation}
\label{f6.1}
\langle X_k\rangle =\frac{\mu}{N_D} +O(\lambda^2/N_D), \qquad {\rm Var}(X_k) = \frac{\mu}{N_D}\big(1- \frac{\mu}{N_D}\big) +O(\lambda^2/N_D),
\end{equation}
where $\mu$ is independent of $k$ and $N_D$, given by
\begin{eqnarray}
\mu =\mu(t) &=&
1 -(1-e^{\i t\varepsilon_1^{(2)}}) \frac{1-\alpha^2}{1+\alpha^2}\ \Big[ \frac{1}{1+\alpha^2} -\frac{e^{-\beta e_1}}{e^{-\beta e_1}+e^{-\beta e_2}} \Big]\nonumber \\
&& - \frac{2\alpha^2}{(1+\alpha^2)^2}\, \big[1- e^{-t {\rm Im} \varepsilon_1^{(3)}} \cos\big(t \, {\rm Re} \,\varepsilon_1^{(3)} \big) \big].  
\label{f2.1}
\end{eqnarray}
We conclude from \eqref{f6.1} that 
\begin{equation}
\label{f4}
\langle X_k\rangle =O(1/N_D)\qquad \mbox{and}\qquad {\rm Var}(X_k) = O\big(1/N_D\big).
\end{equation}
On the other hand, we have from \eqref{f14}, \eqref{f11.1} and the value of $\langle X_k\rangle$ given in \eqref{f6.1} that
\begin{equation}
\label{f10}
{\rm Var}(F_{N_D}) = \frac{\mu(1-\mu)}{(N_D)^2}+O\big(\tfrac{\lambda^2}{(N_D)^2}\big) = O\big(1/(N_D)^2\big).
\end{equation}
We conclude from \eqref{f4} and \eqref{f10}  that the single level variance is by a factor $N_D$ larger than that of the fluctuation of the average level (measured by $F_{N_D}$).

\subsection{Quasi-degenerate system, broken symmetry}
\label{nosym}

Instead of \eqref{m151.9}, \eqref{m152}, one may consider a system where the symmetry is broken,
\begin{eqnarray}
H_\s &=& \sum_{j=1}^{N_D} (E_D+\varepsilon_j) |D_j\rangle\langle D_j| + \sum_{k=1}^{N_A}(E_A+\eta_k) |A_k \rangle\langle A_k|\nonumber\\
&& +\sum_{j,k}(V+\nu_{j,k}) \big( |A_k\rangle \langle D_j| + |D_j\rangle \langle A_k| \big),\label{m}\\
G &=& \sum_{j=1}^{N_D} (g_D+\gamma_{D,j}) |D_j\rangle\langle D_j| + \sum_{k=1}^{N_A} (g_A+\gamma_{A,k}) |A_k \rangle\langle A_k|, 
\label{mm}
\end{eqnarray}
where $\varepsilon_j, \eta_k, \nu_{jk}, \gamma_{D,j}, \gamma_{A,k}$ measure the deviation from the symmetric situation. For simplicity of the discussion, consider $\nu_{j,k}=\gamma_{D,j}=\gamma_{A,k}=0$, so that the non-symmetric characteristics are determined entirely by $\varepsilon_j$, $\eta_k$, defining the donor and acceptor energy bands of size
\begin{equation}
\delta_D = \max\{|\varepsilon_j-\varepsilon_k|\},\qquad \delta_A = \max\{|\eta_j-\eta_k|\}.
\end{equation}
The method of analysis used here can be extended to the regime
\begin{equation}
\label{regime}
\delta\equiv \max\{\delta_D, \delta_A\} <\!\!< \lambda^2 <\!\!<  |e_1-e_2| .
\end{equation}
The first constraint in \eqref{regime} is called the {\em narrow band regime}. We note that the second inequality in \eqref{regime} is not necessary for our method to work. Indeed, in \cite{Dimer} we dealt with systems where $\lambda$ is not constrained (including strong coupling). Let us for simplicity continue with the discussion in the regime \eqref{regime}. One can carry out the spectral analysis of the resonances of the system in terms of a {\em perturbation theory in the two parameters $\delta$ and $\lambda$}, namely, $\delta/|e_1-e_2|<\!\!<1$, $\lambda^2/|e_1-e_2|<\!\!<1$ and  also  $\delta/\lambda^2<\!\!<1$. 

For a simplified model with one donor and two acceptor levels, we have done a detailed analysis in \cite{MBS}. In the general case considered here, we conjecture the same effects to hold. Namely, there emerge {\em two time scales}, 
$$
t_1\propto \lambda^{-2} \quad \mbox{and}\quad t_2\propto \lambda^2/\delta^2 \propto t_1 (\lambda^2/\delta)^2
$$
satisfying
$$
t_1 <\!\!< t_2.
$$

\begin{itemize}
\item For {\em short times}, $t < t_2$, the system dynamics feels the energy spread $\delta>0$ only as an $O(\delta)$ correction, whereas the interaction with the reservoir already drives irreversible dynamics. In particular, at $t\approx t_1$, the system state has already decayed to a quasi-stationary state which depends on the initial state. This quasi-stationary state is, modulo $O(\delta)$, the final state as predicted by the dynamics with $\delta=0$.

\item For {\em intermediate times} $t_1<t<t_2$, the system state moves away from the manifold of quasi-stationary states (in a well prescribed way, with decay directions and speeds given by resonance theory) and 

\item For {\em large times} $t>t_2$, the system approaches a unique final state, which is the coupled DA-reservoir equilibrium reduced to the DA part.
\end{itemize}

\bigskip

\centerline{\includegraphics[width=12cm, angle=0]{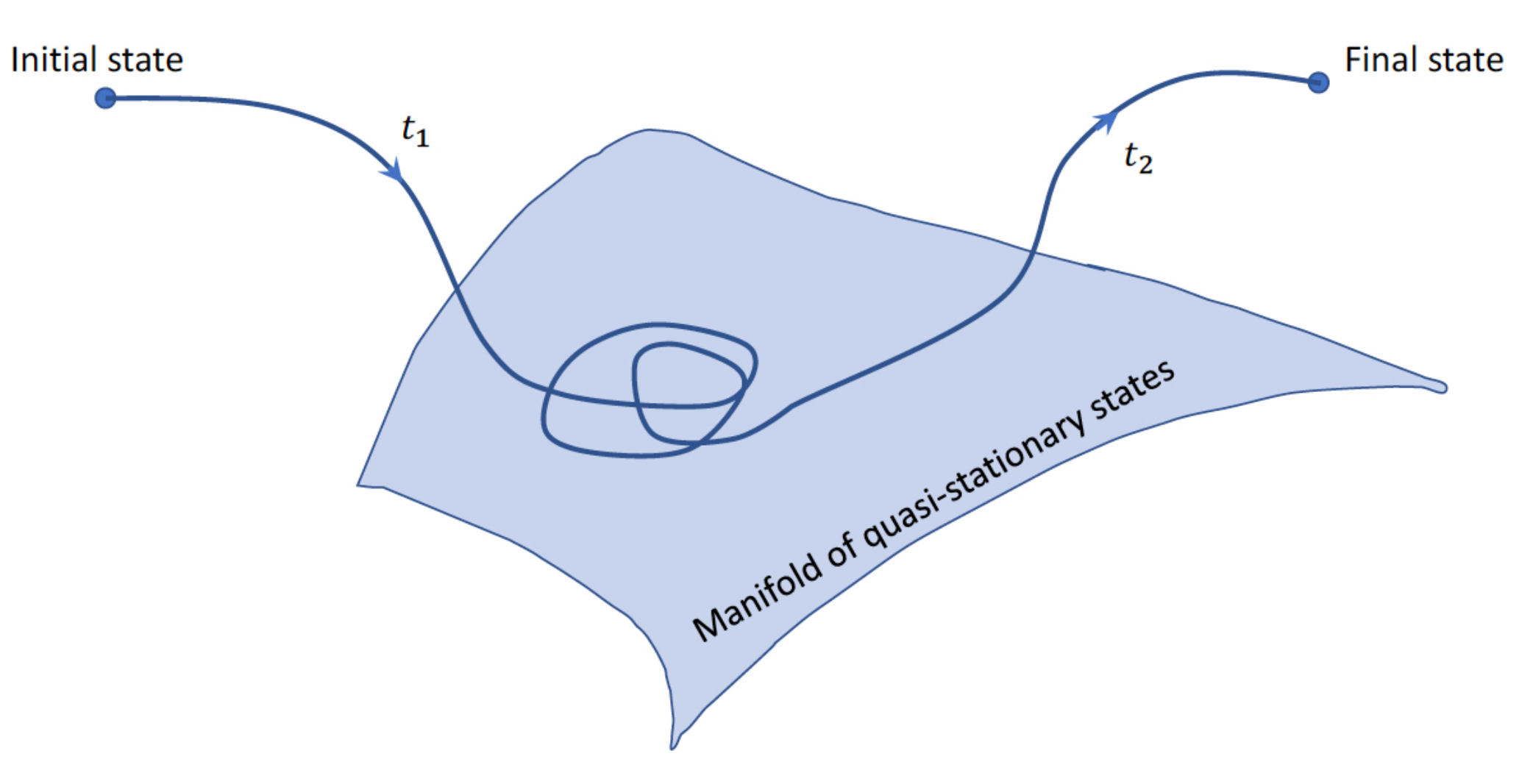}}

\medskip
\centerline{Fig.3:\ The two time-scales $t_1$ and $t_2$.}
\bigskip
\bigskip

Of course, for the convergence to a unique final state to happen for $t>t_2$, one assumes that the coupled system, for $\delta>0$ and $\lambda \neq 0$, has a {\em unique} stationary state (namely, the equilibrium state). This is a condition on the interaction which is generically satisfied for systems without symmetries (called the Fermi Golden Rule Condition), but is {\em not} satisfied in the presence of symmetries.

The above picture is also observed in \cite{GNB}, where a DA model with general $N_A$ and $N_D=1$,  is subjected to a classical noise. It is found there that the dynamics of the acceptor population has two time scales. On the first one, linked to the properties of the classical noise,  the acceptor population approaches the  value $1/2$. This coincides with the value our model gives here (the degenerate case and at high temperature), as explained at the end of Section \ref{secteff}. Then, on the second time scale $\tau_{N_A}\propto (N_A/\delta)^2$, in \cite{GNB} one finds equal population of all levels, hence a total acceptor population of  $1/(N_A+1)$ (since $N_D=1$). This corresponds to the thermal equilibrium value at very high temperature and is consistent with our prediction. The dependence of our $t_2$ (their $\tau_{N_A}$) on $N_D$ and $N_A$ will be revealed by calculating the resonances by perturbation theory in $\delta$ (small). This approach will work rigorously. However, whether we can obtain rigorous bounds on errors valid for all values of $N_D$ and $N_A$ in this perturbation theory remains to be seen. This might be a technical challenge. In the degenerate energy setup considered in the present paper, this difficulty is dodged since the dynamics is reduced to invariant subspaces as a consequence of the symmetry of the Hamiltonian.

\section{Spectrum, resonances of the Liouville operator}

The operator generating the full dynamics is the Liouville operator $L$. It is the equivalent of the Hamiltonian $H$, but expressed in a different Hilbert space than that for $H$. Namely, $L$ acts on the purification Hilbert space of the initial state  $\rho_0=\rho_\s\otimes\rho_{\r,\beta}$, where $\rho_\s$ is an arbitrary DA initial density matrix and $\rho_{\r,\beta}$ is the thermal equilibrium of the reservoir.

\subsection{The purification Hilbert space $\mathcal H$}

We refer to \cite{MSB, Mdyn, Mlnotes} for a detailed exposition of the material in this section. Let $\rho_0$ be an arbitrary density matrix on the DA system, acting on the Hilbert space 
\begin{equation}
{\mathcal H}_\s = {\mathbb C}^{N_D}\otimes {\mathbb C}^{N_A}.
\label{m3} 
\end{equation}
The purification of $\rho_0$ is given by a normalized vector $\Psi_\s$ in the `doubled Hilbert space' ${\mathcal H}_\s\otimes{\mathcal H}_\s$, satisfying ${\rm Tr}(\rho_0 X) = \scalprod{\Psi_\s}{(X\otimes\bbbone_\s)\Psi_\s}$ for any DA observable $X\in{\mathcal B}({\mathcal H}_\s)$. For example, the explicit purification of the DA equilibrium state is given in \eqref{zbetahs}.

Similarly, a well known purification of the equilibrium state of the quantum field of oscillators is given by the so-called {\em Araki-Woods representation} of the canonical commutation relations \cite{AW, Mlnotes}. It is given as follows. For a `single particle wave function' $f\in L^2({\mathbb R}^3, d^3k)$, define $f_\beta\in L^2({\mathbb R}\times S^2, du d\Sigma)$ by
\begin{equation}
f_\beta(u,\Sigma) = \sqrt{\frac{u}{1-e^{-\beta u}}}\, |u|^{1/2} 
\left\{
\begin{array}{ll}
f(u,\Sigma), & u\ge0,\\
-\overline f(-u,\Sigma) & u<0.
\end{array}	
\right.
\label{aw}
\end{equation}
$(u,\Sigma)\in {\mathbb R}\times S^2$ and $f$ on the right side is expressed in spherical coordinates $\vec k =(\omega,\Sigma)\in {\mathbb R}_+\times S^2$. The Araki-Woods, or {\em thermal representation} of the creation and annihilation operators $a^*(f)$, $a(f)$ is given by $a^*(f_\beta)$, $a(f_\beta)$, which are simply the creation and annihilation operators acting on the Fock space
\begin{equation}
{\mathcal F} = {\mathbb C}\otimes \bigoplus_{n\ge 1} L^2({\mathfrak h}_\beta^{\otimes_{\rm s}n}),\qquad {\mathfrak h}_\beta = L^2({\mathbb R}\times S^2, du d\Sigma).
\label{f}
\end{equation}
In \eqref{f}, ${\mathfrak h}_\beta^{\otimes_{\rm s}n}$ is the symmetrized $n$-fold tensor product (Bosons). This is the purification of the reservoir equilibrium state $\rho_{\r,\beta}$ (in the continuous mode, or thermodynamic limit). More precisely, if $P(f_1,\ldots,f_k)$ is any polynomial of $a^*(f_i)$ and $a(f_j)$, then 
$$
{\rm tr}_{\r} \big(\rho_{\r,\beta} P(f_1,\ldots,f_n)\big) = \scalprod{\Omega_\r}{P( (f_1)_\beta,\ldots (f_k)_\beta) \Omega_\r},
$$
where $\Omega_\r$ is the vacuum vector in $\mathcal F$. Accordingly, the thermal field and Weyl operators are defined by
$$
\varphi(f_\beta) = \frac{1}{\sqrt 2}\big(a^*(f_\beta) + a(f_\beta)\big),\qquad W(f_\beta) = e^{i\varphi(f_\beta)}, 
$$
all acting on $\mathcal F$. The total DA-reservoir Liouville (purification) Hilbert space is given by
\begin{equation}
\label{m9}
{\cal H} = {\cal H}_\s\otimes{\cal H}_\s  \otimes  {\cal F}.
\end{equation}

Let $X$ be a DA observable (an operator acting on ${\mathcal H}_\s$). The average of $X$ at time $t$ is given by 
\begin{equation}
\av{X}_t= {\rm Tr}_{\s+\r} \Big( (\rho_0\otimes\rho_{\r,\beta})\, e^{\i tH}(X\otimes\bbbone_\r) e^{-\i tH}\Big)  = \scalprod{\Psi_0}{e^{\i tL} (X\otimes\bbbone_\s\otimes\bbbone_\r)e^{-\i tL}\Psi_0},
\label{m19'}
\end{equation}
where $\Psi_0=\Psi_\s\otimes \Omega_\r$ purifies $\rho_0\otimes\rho_{\r,\beta}$ and where $L$ is the Liouville operator, constructed from the Hamiltonian \eqref{hamiltonian}, given by
\begin{eqnarray}
\label{m11}
L &=& L_\s +L_\r +\lambda I,\nonumber\\
L_\s &=& H_\s\otimes {\bbbone}_\s\otimes\bbbone_{\cal F} - {\bbbone}_\s \otimes H_\s\otimes \bbbone_{\cal F},\nonumber\\
I&=& G\otimes\bbbone_\s\otimes \varphi(h) - \bbbone_\s\otimes G\otimes\widetilde\varphi(h).
\end{eqnarray}
The $\bbbone_\s$ is the identity operator on ${\mathcal H}_\s$.
The free field Liouvillian, given by
\begin{equation}
L_\r = \int_{{\mathbb R}\times S^2} u\  a^*(u,\Sigma) a(u,\Sigma) d u d\Sigma,
\end{equation}
is self-adjoint for any value of $\lambda\in\mathbb R$.  This is proven for instance by using {\em Glimm-Jaffe-Nelson triples} techniques, {\em c.f.} \cite{FM}. 

In \eqref{m11} and in what follows, we simply write $\varphi(h)$ instead of $\varphi(f_\beta)$ for the thermal field operator and we have introduced $\widetilde\varphi(h):=\widetilde\varphi(\bar h_\beta(-u,\Sigma))$. This quantity is related to the {\em modular conjugation} $J_\r$ of the thermal field (see {\em e.g.} \cite{AW,MSB} and also \eqref{m114}), defined by
\begin{equation}
\label{jr}
J_\r \varphi(f_\beta(u,\Sigma)) J_\r = \widetilde\varphi(\bar f_\beta(-u,\Sigma)).
\end{equation}

The equilibrium state with respect to the interacting dynamics is represented in the purification Hilbert space by the `interacting KMS vector' (Kubo-Martin-Schwinger) \cite{MSB}
\begin{equation}
\label{m34}
\Omega_{\s\r,\beta,\lambda} = \frac{e^{-\beta (L_0+\lambda G\otimes\bbbone_\s\otimes\varphi(h))/2} \Omega_{\s,\beta}\otimes \Omega_\r}{\| e^{-\beta (L_0+\lambda G\otimes\bbbone_\s\otimes\varphi(h))/2} \Omega_{\s,\beta}\otimes \Omega_\r\| },
\end{equation}
where (c.f. \eqref{m11})
\begin{equation}
\label{m35}
L_0=L_\s+L_\r.
\end{equation}
Here $\Omega_{\s,\beta}$ and $\Omega_\r$ are the (purifications) of the system and reservoir equilibrium (KMS) states.  $\Omega_\r$ is simply the vaccum vector in $\mathcal F$, \eqref{f} and the explicit form of $\Omega_{\s,\beta}$ is given in \eqref{zbetahs}.

\subsection{Decomposition into invariant subspaces}

We introduce the decomposition
\begin{equation}
\label{m8}
{\mathcal H}_\s = \bar{\cal H}_\s\oplus \bar{\cal H}_\s^\perp,\qquad 
\bar{\cal H}_\s={\rm span}\{|D\rangle, |A\rangle\},
\end{equation}
where $\bar{\cal H}_\s$ is the {\em effective two-level DA space} and we set $\bar{\cal H}_\s^\perp\equiv (\bar{\cal H}_\s)^\perp$. 
The operators $H_\s$ and $G$, \eqref{m152}, are reduced (block diagonal) in this decomposition,
\begin{equation}
\label{m3}
H_\s = \bar H_\s\oplus \bar H_\s^\perp,\quad G = \bar G\oplus \bar G^\perp,
\end{equation}
written in block form as
\begin{eqnarray}
\label{m4}
\bar H_\s = 
\begin{pmatrix}
E_D & v\\
v & E_A
\end{pmatrix} ,\quad & \quad
\bar H_\s^\perp = 
\left(\begin{array}{@{}c|c@{}}
E_D{\mathbf 1} & {\mathbf 0} \\ \hline
{\mathbf 0} & E_A{\mathbf 1}
\end{array}\right),\nonumber\\
\bar G = 
\begin{pmatrix}
g_D & 0\\
0& g_A
\end{pmatrix}, \quad & \quad  \bar G^\perp = \left(\begin{array}{@{}c|c@{}}
g_D {\mathbf 1} & {\mathbf 0} \\ \hline
{\mathbf 0} & g_A {\mathbf 1}
\end{array}\right).
\end{eqnarray}
The diagonalization of $\bar H_\s$ is given by $\bar H_\s = \sum_{j=1,2} e_j |\varphi_j\rangle\langle\varphi_j|$, c.f. \eqref{m22}, \eqref{e12'}. The decomposition \eqref{m8} induces a decomposition of the total Liouville Hilbert space \eqref{m9} into four parts,
\begin{equation}
\label{m10}
{\cal H} = \bigoplus_{j=1}^4 {\cal H}_j , \qquad \mbox{with}
\qquad\qquad 
\begin{array}{lll}
{\cal H}_1&=&\bar{\cal H}_\s\otimes \bar{\cal H}_\s\otimes{\cal F},\\
{\cal H}_2&=&\bar{\cal H}_\s\otimes \bar{\cal H}_\s^\perp \otimes{\cal F},\\
{\cal H}_3&=&\bar{\cal H}_\s^\perp \otimes \bar{\cal H}_\s\otimes{\cal F},\\
{\cal H}_4&=&\bar{\cal H}_\s^\perp\otimes \bar{\cal H}_\s^\perp\otimes{\cal F}.
\end{array}
\end{equation} 
The Liouvillian $L$, \eqref{m11}, is block diagonal in this direct sum decomposition,
\begin{equation}
\label{m10.1}
L = \bigoplus_{j=1}^4 L_j
\end{equation}
with
\begin{eqnarray}
L_1&=&\bar H_\s\otimes \bbbone_\s - \bbbone_\s\otimes \bar H_\s +L_\r + \lambda \bar G\otimes\bbbone_\s\otimes\varphi(h) - \lambda\bbbone_\s\otimes \bar G\otimes\widetilde\varphi(h),\label{L1}\\
L_2&=&\bar H_\s\otimes \bbbone_\s - \bbbone_\s\otimes \bar H_\s^\perp +L_\r + \lambda \bar G\otimes\bbbone_\s\otimes\varphi(h) -\lambda \bbbone_\s\otimes \bar G^\perp\otimes\widetilde\varphi(h),\label{L2}\\
L_3&=&\bar H_\s^\perp\otimes \bbbone_\s - \bbbone_\s\otimes \bar H_\s +L_\r + \lambda \bar G^\perp\otimes\bbbone_\s\otimes\varphi(h) - \lambda\bbbone_\s\otimes \bar G\otimes\widetilde\varphi(h),\label{L3}\\
L_4&=&\bar H_\s^\perp\otimes \bbbone_\s - \bbbone_\s\otimes \bar H_\s^\perp +L_\r + \lambda \bar G^\perp\otimes\bbbone_\s\otimes\varphi(h) -\lambda \bbbone_\s\otimes \bar G^\perp\otimes\widetilde\varphi(h).\quad \label{L4}
\end{eqnarray}
The various $\bbbone_\s$ in \eqref{L1}-\eqref{L4} are understood to be the identity operators on the appropriate (sub-)spaces.

\subsection{Spectral analysis of $L$}

Due to the decomposition \eqref{m10.1}, \eqref{L1}-\eqref{L4}, the spectrum of $L$ is the union
\begin{equation}
{\rm spec}(L) = \bigcup_{j=1}^4 {\rm spec} (L_j).
\end{equation}

\subsubsection{Spectrum and resonances of $L_1$}

The operator $L_1$ describes the dynamics of an {\em effective spin-boson} system, that is, a two-level system coupled to a bosonic reservoir. The dynamics of this system is not explicitly solvable, but can be analyzed by perturbation theory in the setting \eqref{m20'}.\footnote{We note that it is also of interest to consider the {\em strong DA-reservoir interaction regime}, characterized by $v <\!\!< \lambda^2 <\!\!< E_D-E_A$. This regime may be treated by first solving the problem for $v=0$ and then implementing a perturbation theory for small $v$, as has been done in \cite{Dimer}.}

The purification of the the effective DA equilibrium state \eqref{m142} is
\begin{equation}
\label{m64}
\bar\Omega_{\s,\beta} =\bar Z_{\s,\beta}^{-1/2} \Big(e^{-\beta e_1/2}\varphi_1\otimes\varphi_1 +e^{-\beta e_2/2}\varphi_2\otimes\varphi_2\Big)\in \bar{\cal H}_\s\otimes\bar {\cal H}_\s, \ \bar Z_{\s,\beta}=e^{-\beta e_1} +e^{-\beta e_2}.
\end{equation}
When coupled to the reservoir, the effective two level system has an {\em effective coupled equilibrium state}, which is given according to the perturbation theory of KMS (equilibrium) states by the vector (in the purification Hilbert space)
\begin{equation}
\label{m63}
\bar \Omega_{\s\r,\beta,\lambda} = Z_{\beta,\lambda}^{-1/2} e^{-\beta (\bar H_\s\otimes\bbbone_\s - \bbbone_\s \otimes \bar H_\s+ L_\r+\lambda \bar G\otimes\bbbone_\s\otimes\varphi(h))/2}\big( \bar\Omega_{\s,\beta}\otimes \Omega_\r \big) \in \bar{\cal H}_\s\otimes \bar{\cal H}_\s\otimes{\cal F},
\end{equation}
where $\Omega_\r$ is the vacuum in $\cal F$ and $Z_{\beta,\lambda}^{-1/2}$ is a normalization factor. By the construction of the Liouville operator, we know a priori that \begin{equation}
L_1\bar\Omega_{\s\r,\beta,\lambda}=0.
\end{equation}

The eigenvalues of $\bar H_\s$ are $e_1$ and $e_2$, given in \eqref{e12'}, with associated eigenvectors $\varphi_{1,2}$, \eqref{m22}. The nonzero eigenvalues of  $\bar L_0= \bar H_\s\otimes\bbbone - \bbbone\otimes \bar H_\s +L_\r$, acting on $\bar{\cal H}_\s\otimes\bar{\cal H}_\s\otimes{\cal F}$, are the simple eigenvalues $\pm(e_1-e_2)$ with associated eigenvectors $\varphi_{12}\otimes\Omega_\r$ and $\varphi_{21}\otimes\Omega_\r$, where  $\varphi_{12}=\varphi_1\otimes\varphi_2$, $\varphi_{21}= \varphi_2\otimes\varphi_1$ and $\Omega_\r$ is the vacuum in $\cal F$. The other eigenvalue of $\bar L_0$ is zero and is doubly degenerate with eigenvectors $\varphi_{11}\otimes \Omega_\r$ and $\varphi_{22}\otimes\Omega_\r$, with $\varphi_{11}=\varphi_1\otimes\varphi_1$ and $\varphi_{22}=\varphi_2\otimes\varphi_2$. These eigenvalues are embedded in continuous spectrum covering all of $\mathbb R$. As the interaction $\lambda  \bar G\otimes\bbbone_\s\otimes\varphi(h) - \lambda\bbbone_\s\otimes\bar G\otimes\widetilde \varphi(h)$ is switched on, the above eigenvalues of $\bar L_0$ become {\em resonances}, some of them acquiring non-vanishing imaginary parts. To describe them, introduce $[\bar G]_{ij}=\scalprod{\varphi_i}{\bar G\varphi_j}$. A calculation gives
\begin{eqnarray}
[\bar G]_{ij} &=& [\bar G](e_i,e_j), \nonumber\\
{}[\bar G](a,b) &=& \frac{g_Dv^2 +g_A(a-E_D)(b-E_D)}{\sqrt{\big(v^2+(a-E_D)^2\big) \big( v^2+(b-E_D)^2\big)} }.
\label{gij}
\end{eqnarray}

\begin{lem}
\label{lemma2}
The resonance energies and eigenvectors of $L_1$ are as follows.
\begin{itemize}
\item[{\bf(A)}] Zero remains an eigenvalue of $L_1$ also for $\lambda\neq 0$. Call it $\varepsilon_1^{(1)}=0$. The associated eigenvector is the vector $\bar\Omega_{\s\r,\beta,\lambda}$ (c.f. \eqref{m63}), which has the expansion
\begin{eqnarray}
\bar\Omega_{\s\r,\beta,\lambda}&=& \bar\Omega_{\s,\beta}\otimes\Omega_\r +{\mathcal O}(\lambda^2),\nonumber\\
\bar\Omega_{\s,\beta} &=&  \frac{e^{-\beta e_1/2}\varphi_{11} +e^{-\beta e_2/2}\varphi_{22}}{\sqrt{{\rm Tr}e^{-\beta \bar H_\s}}},\qquad {\rm Tr} e^{-\beta \bar H_\s} = e^{-\beta e_1} +e^{-\beta e_2}.
\label{m46}
\end{eqnarray}
The other resonance close to the origin is $\varepsilon_1^{(2)} = \i\lambda^2(  \gamma_0+O(\lambda^2))$, with associated resonance eigenvector $\Psi_1^{(2)}\otimes\Omega_\r +{\mathcal O}(\lambda^2)$, where
\begin{eqnarray}
\gamma_0 &=&  \frac{8}{\beta} \big( [\bar G]_{11}^2+[\bar G]_{22}^2 \big) \widetilde J(0) +4[\bar G]^2_{12} \coth\big(\beta |e_1-e_2|/2\big) \ J(|e_1-e_2|), \nonumber\\
\Psi_1^{(2)}&=& \frac{ e^{\beta e_1/2}\varphi_{11} - e^{\beta e_2/2}\varphi_{22}}{\sqrt{e^{\beta e_1} +e^{\beta e_2} }}.
\label{m47}
\end{eqnarray}  
\item[{\bf (B)}] A resonance bifurcates out of $e_1-e_2$. It has energy 
$$
\varepsilon_1^{(3)} = e_1-e_2+\lambda^2( x_{12}+\i  y_{12})+O(\lambda^4)
$$
and eigenvector $\varphi_{12}\otimes\Omega_\r +{\mathcal O}(\lambda^2)$, with
\begin{eqnarray}
y_{12} &=& \frac{2}{\beta}  \big( [\bar G]^2_{11} + [\bar G]^2_{22} + 2[\bar G]_{12}^2\big)\widetilde J(0)  +2[\bar G]^2_{12}\coth\big(\beta|e_1-e_2|/2\big)\ J(|e_1-e_2|), \nonumber\\
x_{12} &=& -\frac12\big( [\bar G]_{11}^2-[\bar G]_{22}^2 \big )\  {\rm P.V.} \int_{{\mathbb R}\times S^2} |g_\beta(u,\Sigma)|^2 \frac 1u du\Sigma\nonumber\\
&& -\frac12 [\bar G]_{12}^2 \ {\rm P.V.} \int_{{\mathbb R}\times S^2} |g_\beta(u,\Sigma)|^2\Big\{ \frac{e^{-\beta u}}u +\frac{1}{u+e_2-e_1} -\frac{1}{u+e_1-e_2}   \Big\} du d\Sigma.\nonumber\\
\label{m23}
\end{eqnarray}
\item[{\bf (C)}] A resonance bifurcates out of $e_2-e_1$. It has energy
\begin{equation}
\varepsilon_1^{(4)} = e_2-e_1+\lambda^2( -x_{12}+\i  y_{12})+O(\lambda^4)
\label{m23.1}
\end{equation}
with resonance eigenstate $\varphi_{21}\otimes\Omega_\r +{\mathcal O}(\lambda^2)$. The $x_{12}$, $y_{12}$ of \eqref{m23.1} are given in \eqref{m23}.
\end{itemize}
\end{lem}

{\bf Proof of Lemma \ref{lemma2}.\ } The arguments are standard within resonance theory, see \cite{MSB, Mlso, Mdyn}. We explain them without giving the explicit calculations. 

{\bf (A)} The state $\bar\Omega_{\s,\beta}\otimes\Omega_\r$ is a KMS state (equilibrium at temperature $1/\beta$) w.r.t. the dynamics generated by $L_1|_{\lambda=0}$.  By perturbation theory of KMS (equilibrium) states, we know that $\bar\Omega_{\s\r,\beta,\lambda}$ is a KMS state for $L_1$ and hence $L_1\bar\Omega_{\s\r,\beta,\lambda}=0$ also for $\lambda\neq 0$. This shows that the eigenvalue at the origin persists under perturbation. To track the fate of the remaining part of the twofold degeneracy of zero as an eigenvalue of $L_0$ (restricted to ${\mathcal H}_1$), we utilize perturbation theory. Namely, the resonance energies and eigenvectors are obtained from the {\em level shift operator} 
\begin{equation}
\Lambda_0 = -\lambda^2 P_0 I \bar P_0 (L_0+\i 0_+)^{-1} \bar P_0 I P_0,
\label{n1}
\end{equation}
where $P_0$ is the eigenprojection of $L_0$ associated with $e=0$ and $\bar P_0=\bbbone-P_0$. Here, the perturbation (interaction) operator is (c.f. \eqref{L1}) $I= 
 \bar G\otimes\bbbone_\s\otimes\varphi(h) - \bbbone_\s\otimes \bar G\otimes\widetilde\varphi(h)$. 
The operator $\Lambda_0$ is identified as a two-by-two matrix acting on ${\rm span}\{\varphi_1,\varphi_2\}$. A standard calculation gives 
\begin{eqnarray}
\scalprod{\varphi_1}{\Lambda_0 \varphi_1} &=& 4\i \Big\{ \frac{2}{\beta} [\bar G]_{11}^2\widetilde J(0) + [\bar G]_{12}^2\frac{J(|e_1-e_2|)}{|1-e^{-\beta(e_1-e_2)}|}\Big\}, \nonumber\\
\scalprod{\varphi_2}{\Lambda_0 \varphi_2} &=& 4\i \Big\{ \frac{2}{\beta} [\bar G]_{22}^2\widetilde J(0) + [\bar G]_{12}^2\frac{J(|e_1-e_2|)}{|1-e^{+\beta(e_1-e_2)}|}\Big\}.
\label{n2}
\end{eqnarray}
We know {\em a priori} that $\Lambda_0\bar\Omega_{\s,\beta}=0$ (which follows from $L_1\bar\Omega_{\s\r,\beta,\lambda}=0$) and so the second eigenvalue of $\Lambda_0$ is its trace, namely, the sum of the two terms in \eqref{n2}, which is  $\i\lambda^2 \gamma_0$ with $\gamma_0$ given in \eqref{m47}. The eigenvector associated to this nonzero eigenvalue has to be orthogonal to $\Omega_{\s,\beta}$ and hence it is as in \eqref{m47}.  

{\bf (B)} The level shift operator associated to the simple eigenvalue $e_1-e_2$ of $L_0$ is one-dimensional, 
\begin{equation}
\Lambda_{12} = -\lambda^2 \scalprod{\varphi_{12}\otimes\Omega_\r}{ I(L_0+\i 0_+)^{-1}I \varphi_{12}\otimes\Omega_\r} P_{12}
= \lambda^2 (x_{12}+\i y_{12}) P_{12},
\label{n4}
\end{equation}
where $P_{12} = |\varphi_{12}\otimes\Omega_\r\rangle \langle \varphi_{12}\otimes\Omega_\r|$ and $x_{12}$, $y_{12}$ are given in \eqref{m23}.

{\bf (C)} This follows again by calculation, or more elegantly, from the well known symmetry \cite{Mlso} between the level shift operator associated to $e_1-e_2$ and that of $e_2-e_1$. \hfill \qed

\subsubsection{Spectrum and resonances of  $L_2$}
\label{sectL2L3}

Consider the decomposition
\begin{equation}
\label{m12}
(\bar{\cal H}_\s)^\perp = \bar{\cal H}_{D\perp}\oplus \bar{\cal H}_{A\perp}
\end{equation}
where $\bar {\cal H}_{D\perp}$ and $\bar {\cal H}_{A\perp}$ are the subspaces introduced in points 1 and 2 before \eqref{m142}. Recall that $P_{D\perp}$ and $P_{A\perp}$ are the orthogonal projections onto $\bar{\cal H}_{D\perp}$, $\bar{\cal H}_{A\perp}$. The operator $L_2$ is block diagonal in the decomposition 
$$
\bar{\cal H}_\s\otimes \bar{\cal H}^\perp_\s\otimes{\cal F} = \big(\bar{\cal H}_\s\otimes\bar{\cal H}_{D\perp}\otimes {\cal F}\big)  \oplus \big(\bar{\cal H}_\s\otimes\bar{\cal H}_{A\perp}\otimes {\cal F}\big).
$$
 We have
\begin{equation}
\bar H_\s^\perp P_{D\perp} = E_D P_{D\perp},\quad  \bar H_\s^\perp P_{A\perp} = E_A P_{A\perp},\quad  \bar G_\s^\perp P_{D\perp} = g_D P_{D\perp},\quad  \bar G_\s^\perp P_{A\perp} = g_A P_{A\perp}.
\label{therelations}
\end{equation}
The restrictions of $L_2$ to $\bar{\cal H}_\s\otimes\bar{\cal H}_{D\perp}\otimes {\cal F}$ and  $\bar{\cal H}_\s\otimes\bar{\cal H}_{A\perp}\otimes {\cal F}$ are, respectively
\begin{eqnarray}
\label{m24}
L_{2,D\perp} &=& \Big( \bar H_\s - E_D  +L_\r +\lambda \bar G\otimes\varphi(h) -\lambda g_D\widetilde\varphi(h) \Big)\otimes P_{D\perp},\label{m25.1}\\
L_{2,A\perp} &=& \Big( \bar H_\s - E_A  +L_\r +\lambda \bar G\otimes\varphi(h) -\lambda g_A\widetilde\varphi(h) \Big)\otimes P_{A\perp},
\label{m25}
\end{eqnarray}
where the operators in the parentheses act on $\bar{\cal H}_\s\otimes{\cal F}$. The spectrum of $L_2$ is the union of the spectra of $L_{2,D\perp}$ and $L_{2,A\perp}$.

For $\lambda=0$, the eigenvalues of $L_{2,D\perp}$ are $e_1-E_D$ and $e_2-E_D$, both having degeneracy $N_D-1$ ($={\rm rank}P_{D\perp}$). The eigenspaces are ${\rm Ran}\ |\varphi_1\rangle\langle\varphi_1| \otimes P_{D\perp}\otimes |\Omega_\r\rangle\langle\Omega_\r|$ and ${\rm Ran}\ |\varphi_2\rangle\langle\varphi_2| \otimes P_{D\perp}\otimes |\Omega_\r\rangle\langle\Omega_\r|$, respectively (where Ran  stands for the range of a projection). In the same way, for $\lambda=0$, the eigenvalues of $L_{2,A\perp}$ are $e_1-E_A$ and $e_2-E_A$, both having degeneracy $N_A-1$. The eigenspaces are ${\rm Ran}\ |\varphi_1\rangle\langle\varphi_1| \otimes P_{A\perp}\otimes |\Omega_\r\rangle\langle\Omega_\r|$ and ${\rm Ran}\ |\varphi_2\rangle\langle\varphi_2| \otimes P_{A\perp}\otimes |\Omega_\r\rangle\langle\Omega_\r|$, respectively. 

To analyze the spectrum of $L_2$ for $\lambda\neq 0$, we proceed as follows. We perform a polaron transformation to get rid of the terms $-\lambda g_D\widetilde\varphi(h)$ and $-\lambda g_A\widetilde\varphi(h)$ in \eqref{m24} and \eqref{m25}, and then we do perturbation theory in $\lambda$. 

\begin{lem}
\label{lemma3}
All four eigenvalues of $L_2$ for $\lambda=0$ turn into resonances, given by 
\begin{eqnarray}
\varepsilon_2^{(1)} &=& e_1-E_D + \lambda^2 x_1 +2\i\lambda^2\Big( \frac{1}{\beta}[\bar G]^2_{11}\widetilde J(0) +[\bar G]^2_{12}\frac{J(|e_1-e_2|)}{|1-e^{-\beta (e_1-e_2)}|} \Big) +O(\lambda^4),\qquad\nonumber\\
x_1 &=& \frac12 g^2_D\|h_\beta/\sqrt\omega\|^2 -\frac12 [\bar G]_{11}^2\  {\rm P.V.} \int_{{\mathbb R\times S^2}}|h_\beta(u,\Sigma)|^2 \frac1u dud\Sigma \nonumber\\
&& -\frac12  [\bar G]_{12}^2\  {\rm P.V.} \int_{{\mathbb R\times S^2}}|h_\beta(u,\Sigma)|^2 \frac{1}{u-e_1+e_2} dud\Sigma, \label{m26}\\
\varepsilon_2^{(2)} &=& e_2-E_D + \lambda^2 x_2 +2\i\lambda^2\Big( \frac{1}{\beta}[\bar G]^2_{22}\widetilde J(0) +[\bar G]^2_{12}\frac{J(|e_1-e_2|)}{|1-e^{\beta (e_1-e_2)}|} \Big) +O(\lambda^4),\nonumber\\
x_2 &=& \frac12 g^2_D\|h_\beta/\sqrt\omega\|^2 -\frac12 [\bar G]_{22}^2\  {\rm P.V.} \int_{{\mathbb R\times S^2}}|h_\beta(u,\Sigma)|^2 \frac1u dud\Sigma \nonumber\\
&& -\frac12  [\bar G]_{12}^2\  {\rm P.V.} \int_{{\mathbb R\times S^2}}|h_\beta(u,\Sigma)|^2 \frac{1}{u-e_2+e_1} dud\Sigma. \label{m27}
\end{eqnarray}
The other two resonances are $\varepsilon_2^{(3)}$ and $\varepsilon_2^{(4)}$, obtained from the expressions of $\varepsilon_2^{(1)}$ and $ \varepsilon_2^{(2)}$ above in \eqref{m26}, \eqref{m27}  by replacing $E_D$ with $E_A$ and $g_D$ with $g_A$. The  eigenspaces associated to these eigenvalues are 
\begin{eqnarray}
{\cal E}_2^{(1)} &=& {\rm Ran} \Big( |\varphi_1\rangle\langle \varphi_1| \otimes P_{D\perp} \otimes |\Omega_\r\rangle\langle \Omega_\r| \Big)+O(\lambda^2), \label{m30}\\
{\cal E}_2^{(2)} &=& {\rm Ran} \Big( |\varphi_2\rangle\langle \varphi_2| \otimes P_{D\perp} \otimes |\Omega_\r\rangle\langle \Omega_\r| \Big)+O(\lambda^2),\label{m31}\\
{\cal E}_2^{(3)} &=& {\rm Ran} \Big( |\varphi_1\rangle\langle \varphi_1| \otimes P_{A\perp} \otimes |\Omega_\r\rangle\langle \Omega_\r| \Big)+O(\lambda^2),\label{m32}\\
{\cal E}_2^{(4)} &=& {\rm Ran} \Big( |\varphi_2\rangle\langle \varphi_2| \otimes P_{A\perp} \otimes |\Omega_\r\rangle\langle \Omega_\r| \Big)+O(\lambda^2).\label{m33}
\end{eqnarray}
The multiplicity of each distinct resonance is the same as that of the corresponding eigenvalue for $\lambda=0$ (namely, ${\rm rank}P_{D\perp}$ or ${\rm rank}P_{A\perp}$). The $O(\lambda^2)$ terms in \eqref{m30}-\eqref{m33} and the $O(\lambda^4)$ remainders in \eqref{m26}, \eqref{m27} are uniform in $N_D$ and $N_A$. 
\end{lem}

{\em Remark.\ } The exact meaning of ${\rm Ran} \big( |\varphi_1\rangle\langle \varphi_1| \otimes P_{D\perp} \otimes |\Omega_\r\rangle\langle \Omega_\r|\big) +O(\lambda^2)$ in \eqref{m30} (and similarly for  \eqref{m31}-\eqref{m33})  is ${\mathcal E}_2^{(1)} = {\rm Ran} ( |\chi\rangle\langle\chi|\otimes P_{D\perp})$, where $\chi = \varphi_1\otimes\Omega_\r +O(\lambda^2)$ is a vector on the first DA-tensor factor and the reservoir tensor factor. The $O(\lambda^2)$ correction is a vector in ${\cal H}_\s\otimes{\cal F}$ with the $O(\lambda^2)$ property holding uniformly in the size of $N_D$.

\bigskip
{\bf Proof of Lemma \ref{lemma3}.} We first treat $L_{2,D\perp}$. Define the unitary operator $T=J_\r W(\frac{\lambda^2 g_D h}{\i\omega})J_\r$, where $J_\r$ is the reservoir modular conjugation and $W(f)=e^{\i \varphi(f)}$ is the thermal Weyl operator. Note that $T$ commutes with all reservoir observables. We conjugate $L_{2,D\perp}$ with $T$,
\begin{eqnarray}  
T L_{2,D\perp} T^* &=& \big( \bar H_\s +L_\r +\lambda \bar G\otimes\varphi(h) +c\big)\otimes P_{D\perp},
\label{n5}
\end{eqnarray}
where $c = -E_D+\tfrac12 \lambda^2 g_D^2 \|h/\sqrt\omega\|^2$. The operator in parentheses in \eqref{n5} acts on the space ${\mathcal H}_\s\otimes\mathcal F$.  The relation \eqref{n5} is obtained in a standard way by taking into account that for all $f,g$, 
\begin{eqnarray}
W(f) L_\r W(f)^* &=& L_\r -\varphi(\i\omega f) +\tfrac12 \|\sqrt{\omega}f\|^2_2,\nonumber\\
W(f) \varphi(g) W(f)^* &=& \varphi(g)-{\rm Im}\scalprod{f}{g}.
\label{m15}
\end{eqnarray}
We now analyze the spectrum of $\bar H_\s +L_\r +\lambda \bar G\otimes\varphi(h)$. For $\lambda=0$ this operator has the two simple eivenvalues $e_1$ and $e_2$ with associated eigenvectors $\varphi_1\otimes\Omega_\r$ and $\varphi_2\otimes\Omega_\r$. The lowest order correction to $e_1$ due to the perturbation $\lambda \bar G\otimes\varphi(h)$ is (again, given by the level shift operator) $-\lambda^2 \scalprod{\varphi_1\otimes\Omega_\r}{ \big( \bar G\otimes \varphi(h)(\bar H_\s+L_\r -e_1 +\i 0_+)^{-1} \bar G\otimes\varphi(h)\big)  \varphi_1\otimes\Omega_\r}$. An explicit calculation yields \eqref{m26}. The eigenvectors of $TL_{2,D\perp}T^*$ are $\varphi_j\otimes\chi \otimes \Omega_\r +O(\lambda^2)$, for arbitrary normalized $\chi\in{\rm Ran}P_{D\perp}$ and $j=1,2$. Consequently, the ones for $L_{2,D\perp}$ are $T^*\varphi_j\otimes \chi\otimes \Omega_\r +O(\lambda^2)$. But $T^*=\bbbone_\r +O(\lambda^2)$ and hence the eigenvectors of $L_{2,D\perp}$ are of the form $\varphi_j\otimes  \chi\otimes\Omega_\r +O(\lambda^2)$. This proves \eqref{m30}. The results \eqref{m27} and \eqref{m31} are derived in the same way. 

Finally, the same analysis is applicable for the operator $L_{D,A\perp}$, we need just to replace $E_D$ by $E_A$ and $g_D$ by $g_A$ in the final expressions (compare \eqref{m25.1} and \eqref{m25}). \hfill \qed

\subsubsection{Spectrum and resonances of  $L_3$}
\label{sectL3}

Just as for $L_2$, the operator $L_3$ is block diagonal in the decomposition
$$
\bar{\cal H}^\perp_\s\otimes \bar{\cal H}_\s\otimes{\cal F} = \big(\bar{\cal H}_{D\perp}\otimes \bar{\cal H}_\s \otimes {\cal F}\big)  \oplus \big(\bar{\cal H}_{A\perp}\otimes \bar{\cal H}_\s \otimes {\cal F}\big).
$$
The associated blocks are denoted by $L_{3,D\perp}$ and $L_{3,A\perp}$, 
\begin{eqnarray}
\label{m51}
L_{3,D\perp} &=& P_{D\perp}\otimes \Big(E_D  - \bar H_\s +L_\r -\lambda \bar G\otimes\widetilde\varphi(h) +\lambda g_D\varphi(h) \Big),\\
L_{3,A\perp} &=&  P_{A\perp}\otimes \Big( E_A - \bar H_\s +L_\r -\lambda \bar G\otimes\widetilde\varphi(h) +\lambda g_A\varphi(h) \Big).
\label{m52}
\end{eqnarray}
Let ${\mathcal C}$ be the operator of complex conjugation acting on ${\mathcal H}_\s$ (taking complex conjugates of components of vectors when written in the energy basis). Denote by $J_\r$ the reservoir modular conjugation operator. Then ${\mathcal C}\otimes J_\r$ is an antilinear involution (meaning its square is the identity operator) acting on ${\mathcal H}_\s\otimes {\mathcal F}$. We have 
\begin{eqnarray}
	\label{m53}
	\lefteqn{
E_D  - \bar H_\s +L_\r -\lambda \bar G\otimes\widetilde\varphi(h) +\lambda g_D\varphi(h)}\\
&& = -({\mathcal C}\otimes J_\r) \Big( \bar H_\s-E_D +L_\r +\lambda \bar G\otimes\varphi(h) -\lambda g_D\widetilde\varphi(h) \Big) ({\mathcal C}\otimes J_\r).
\nonumber
\end{eqnarray}
It follows from this relation and \eqref{m24}, \eqref{m51} that $z\in {\rm spec}(L_{3,D\perp}) \Leftrightarrow  -\bar z \in {\rm spec}(L_{2,D\perp})$ and that eigenvectors are related by the application of ${\mathcal C}\otimes J_\r$. Analogous relations hold for $L_{3,A\perp}$ and $L_{2,A\perp}$. Consequently, we obtain the following directly from Lemma \ref{lemma3}.
\begin{lem}
	\label{lemma4}
	All four eigenvalues of $L_3$ for $\lambda=0$ turn into resonances, given by
\begin{equation}
\varepsilon_3^{(j)} = -\big( \varepsilon_2^{(j)}\big)^*,\ \ j=1,\ldots,4  \qquad \mbox{(complex conjugate)}
\end{equation} 
where  $\varepsilon_2^{(j)}$ are given in \eqref{m26}-\eqref{m27} (and the  sentence thereafter). 
	The associated eigenspaces  are 
	\begin{eqnarray}
	{\cal E}_3^{(1)} &=& {\rm Ran} \Big( P_{D\perp}\otimes |\varphi_1\rangle\langle \varphi_1|  \otimes |\Omega_\r\rangle\langle \Omega_\r| \Big)+O(\lambda^2), \label{m54}\\
	{\cal E}_3^{(2)} &=& {\rm Ran} \Big( P_{D\perp} \otimes |\varphi_2\rangle\langle \varphi_2| \otimes |\Omega_\r\rangle\langle \Omega_\r| \Big)+O(\lambda^2),\label{m55}\\
	{\cal E}_3^{(3)} &=& {\rm Ran} \Big( P_{A\perp} \otimes |\varphi_1\rangle\langle \varphi_1| \otimes |\Omega_\r\rangle\langle \Omega_\r| \Big)+O(\lambda^2),\label{m56}\\
	{\cal E}_3^{(4)} &=& {\rm Ran} \Big(P_{A\perp} \otimes  |\varphi_2\rangle\langle \varphi_2| \otimes |\Omega_\r\rangle\langle \Omega_\r| \Big)+O(\lambda^2).\label{m57}
	\end{eqnarray}
 The multiplicity of each distinct resonance is the same as that of the corresponding eigenvalues for $\lambda=0$. 
\end{lem}

\subsubsection{Spectrum of $L_4$}

We again use the decomposition \eqref{m12} and the relations \eqref{therelations} to write $L_4$, given in \eqref{L4}, as 
\begin{eqnarray}
L_4 &=& L_\r +\lambda E_D P_{D\perp}\otimes P_{D\perp}\otimes \big(\varphi(h)-\widetilde\varphi(h)\big)\nonumber\\
&&+  (E_D-E_A) P_{D\perp}\otimes P_{A\perp} + L_\r +\lambda P_{D\perp}\otimes P_{A\perp}\otimes \big(E_D\varphi(h)-E_A\widetilde\varphi(h)\big)\nonumber\\
&&+  (E_A-E_D) P_{A\perp}\otimes P_{D\perp} + L_\r +\lambda P_{A\perp}\otimes P_{D\perp}\otimes \big(E_A\varphi(h)-E_D\widetilde\varphi(h)\big)\nonumber\\
&&+  L_\r +\lambda E_A P_{A\perp}\otimes P_{A\perp}\otimes \big(\varphi(h)-\widetilde\varphi(h)\big)\nonumber\\
&=& L_{4,1}+L_{4,2}+L_{4,3}+L_{4,4}.
\end{eqnarray}
Again, the spectrum of $L_4$ is the union of the spectra of  $L_{4,j}$, $j=1,\ldots,4$. 

\begin{lem}
\label{lem1} 
Suppose that the form factor satisfies $\|h_\beta/\sqrt u\|_2<\infty$. The following holds for arbitrary values of the coupling constant $\lambda\in\mathbb R$. 

\smallskip

{\bf (A)} 	 The spectra of $L_{4,1}$ and $L_{4,4}$ consist of an eigenvalue at zero and absolutely continuous spectrum covering $\mathbb R$. We denote these two eigenvalues $\varepsilon_4^{(1)}=\varepsilon_4^{(2)}=0$. The kernels of $L_{4,1}$ and $L_{4,4}$ have dimension $(N_D-1)^2$ and $(N_A-1)^2$, respectively, and are given by 
\begin{eqnarray}
{\rm Ker}L_{4,1} &=& {\rm Ran}\ P_{D\perp}\otimes P_{D\perp}\otimes |\Psi_D\rangle\langle\Psi_D|,\nonumber\\
{\rm Ker}L_{4,4} &=& {\rm Ran}\ P_{A\perp}\otimes P_{A\perp}\otimes |\Psi_A\rangle\langle\Psi_A|,
\label{m59}
\end{eqnarray}
where $\Psi_X$ is given by (recall \eqref{jr} defining $J_\r$)
\begin{equation}
\label{m14}
\Psi_X =
W\Big(\i \lambda E_X h/\omega\Big) J_\r W\Big(\i \lambda E_X h/\omega\Big)  J_\r \Omega_\r \in {\cal F},\qquad X=D,A.
\end{equation}

{\bf (B)} The spectrum of $L_{4,2}$ consists of the eigenvalue \begin{equation}
\label{m16.1}
\varepsilon_4^{(3)}= E_D-E_A -\tfrac12 \lambda^2 (E_D^2-E_A^2) \, \|h/\sqrt\omega\|^2_2
\end{equation}
having multiplicity $(N_D-1)(N_A-1)$ and purely absolutely continuous spectrum covering $\mathbb R$. The eigenspace is given by ${\rm Ran} P_{D\perp}\otimes P_{A\perp}\otimes |\Psi_{DA}\rangle\langle\Psi_{DA}|$, where $\Psi_{DA}$ is given in \eqref{psida}.

{\bf (C)} The spectrum of $L_{4,3}$ consists of the eigenvalue $\varepsilon_4^{(4)} = -\varepsilon_4^{(3)}$ (with $\varepsilon_4^{(3)}$ as given in \eqref{m16.1}) having multiplicity $(N_D-1)(N_A-1)$, and of purely absolutely continuous spectrum covering $\mathbb R$. The eigenspace is given by ${\rm Ran} P_{A\perp}\otimes P_{D\perp}\otimes |\Psi_{AD}\rangle\langle\Psi_{AD}|$, where $\Psi_{AD}$ (and for later use, $\Psi_{DA}$) are given by  
\begin{equation}
\label{psida}
\Psi_{DA} = W\Big(\frac{E_D h}{\i \omega}\Big) J_\r W\Big(\frac{E_A h}{\i \omega}\Big) J_\r\Omega_\r\quad \mbox{and}\quad 
\Psi_{AD} = W\Big(\frac{E_A h}{\i \omega}\Big) J_\r W\Big(\frac{E_D h}{\i \omega}\Big) J_\r\Omega_\r.
\end{equation}

\end{lem}

{\bf Proof.\ }  (A) Consider the operator $L_\r +\alpha (\varphi(h)-\widetilde\varphi(h))$, $\alpha\in\mathbb R$, which acts purely on the reservoir Hilbert space. Here, $\widetilde \varphi(h) = J_\r\varphi(h) J_\r$, see \eqref{jr}.  We introduce the polaron transformation, given by conjugation with the bounded operator
$$
T=W\Big(\frac{\alpha h}{\i \omega}\Big) J_\r W\Big(\frac{\alpha h}{\i \omega}\Big) J_\r,
$$
where $W(f)=e^{\i \varphi(f)}$ is the field operator. The relations \eqref{m15} with $f=\frac{\alpha h}{\i\omega}$ and $g=h$, give 
\begin{equation}
\label{m13}
T\Big( L_\r +\alpha (\varphi(h)-\widetilde\varphi(h))\Big)T^* = L_\r.
\end{equation}
It follows from \eqref{m13} that
\begin{equation}
\label{m14.2}
T L_{4,1} T^* = L_\r \big( P_{D\perp}\otimes P_{D\perp}\otimes {\bbbone}_{\cal F}\big).
\end{equation}
The statement (A) of the lemma follows from the unitary equivalence \eqref{m14.2}. 

Now we prove (B). Consider 
\begin{eqnarray}
L_{4,2} &=& (E_D-E_A)P_{D\perp}\otimes P_{A\perp} +L_\r + P_{D\perp}\otimes P_{A\perp}\otimes \big(\varphi(\lambda E_Dh) -\widetilde\varphi(\lambda E_A h)\big)\nonumber\\
&=& \Big( (E_D-E_A)+L_\r +\varphi(\lambda E_Dh) -\widetilde\varphi(\lambda E_A h)\Big) P_{D\perp}\otimes P_{A\perp}\otimes \bbbone_{\cal F}.
\label{m17}
\end{eqnarray} 
We modify the above polaron transformation $T$ to
$$
T' = W\Big(\frac{\lambda E_D h}{\i \omega}\Big) J_\r W\Big(\frac{\lambda E_A h}{\i \omega}\Big) J_\r.
$$
Then, again using \eqref{m15} one verifies readily that 
\begin{equation}
\label{m16}
T' \Big( L_\r + \varphi(\lambda E_D h)-\widetilde\varphi(\lambda E_A h)\Big)(T')^* = L_\r-\tfrac{\lambda^2}{2} (E_D^2-E_A^2) \, \|h/\sqrt\omega\|^2_2.
\end{equation}
Combining \eqref{m16} with \eqref{m17} we arrive at
\begin{equation}
\label{m18}
T' L_{4,2} (T')^* = \Big( L_\r + (E_D-E_A) -\tfrac{\lambda^2}{2}(E_D^2-E_A^2) \, \|h/\sqrt\omega\|^2_2\Big) P_{D\perp}\otimes P_{A\perp}\otimes \bbbone_{\cal F}.
\end{equation}
The statements in (B) now follow from the unitary equivalence \eqref{m18}. The proof of (C) is entirely the same as that of (B). This completes the proof of Lemma \ref{lem1}. \hfill \qed

\section{The dynamics}

\subsection{Resonance theory}

According to \eqref{m19'}, the average of $X$ at time $t$ is given by 
\begin{equation}
\av{X}_t= \scalprod{\Psi_0}{e^{\i tL} (X\otimes\bbbone_\s\otimes\bbbone_\r)e^{-\i tL}\Psi_0},
\label{m19}
\end{equation}
where $\Psi_0 = \Psi_\s\otimes\Omega_\r$, with $\Psi_\s$ the purification of the initial DA density matrix $\rho_0$ and $\Omega_\r$ the vacuum vector in $\mathcal F$, \eqref{f}. 

We start by giving the purification vector $\Omega_{\s,\beta}$ representing the DA equilibrium density matrix 
\begin{equation}
\label{m36}
\rho_{\s,\beta} = Z^{-1}_{\s,\beta}\  e^{-\beta H_\s} = Z^{-1}_{\s,\beta}\  e^{-\beta (\bar H_\s +\bar H_\s^\perp)} = Z_{\s,\beta}^{-1}\Big( \bar P_\s  e^{-\beta \bar H_\s} + \bar P_\s^\perp e^{-\beta \bar H_\s^\perp}\Big),
\end{equation}
where $\bar P_\s =|D\rangle\langle D| +|A\rangle\langle A|$ is the projection onto $\bar {\cal H}_\s$, see \eqref{m8}. The eigenvalues $e_{1,2}$ and eigenvectors $\varphi_{1,2}$ of $\bar H_\s$ are given in \eqref{e12'} and \eqref{m22}. To express the purification of $\rho_{\s,\beta}$ as a normalized vector in ${\cal H}_\s\otimes{\cal H}_\s$, we introduce
\begin{equation}
\label{m37}
\{ \xi_{D,1},\ldots,\xi_{D,N_D-1}\} \quad\mbox{and}\quad  \{ \xi_{A,1},\ldots,\xi_{A,N_A-1}\},
\end{equation}
which are orthonormal bases of ${\rm Ran} P_{D\perp}$ and ${\rm Ran} P_{A\perp}$, respectively (see after \eqref{m158}). Each $\xi_{D,j}$ is a vector in ${\mathbb C}^{N_D+N_A}$ whose last $N_A$ components are zero and whose first $N_D$ components add up to zero ($\xi_{D,j}\perp |D\rangle$). Each $\xi_{A,j}$ has vanishing first $N_D$ components and the sum of the other ones is zero. The union of all $\xi_{D,j}$ and $\xi_{A,j}$ is an orthonormal basis of ${\rm Ran}(\bar P_\s)^\perp $.  The purification of the density matrix $\rho_{\s,\beta}$ is the vector $\Omega_{\s,\beta}\in {\mathcal H}_\s\otimes{\mathcal H}_\s$ given by
\begin{eqnarray}
\Omega_{\s,\beta} &=& Z_{\s,\beta}^{-1/2} \Big( \sum_{j=1,2} e^{-\beta e_j/2} \varphi_j\otimes\varphi_j + \sum_{\alpha=A,D}e^{-\beta E_\alpha/2} \sum_{j=1}^{N_\alpha-1}  \xi_{\alpha,j}\otimes\xi_{\alpha,j} \Big), \label{m38}\\
Z_{\s,\beta} &=& {\rm Tr} e^{-\beta H_\s}  =  \sum_{j=1,2} e^{-\beta e_j} +\sum_{\alpha=1,2} (N_\alpha-1) e^{-\beta E_\alpha}. 
\label{zbetahs}
\end{eqnarray}
The  interacting equilibrium state \eqref{m34} is {\em separating} (a property following from the general theory of KMS equilibrium states \cite{BR}), meaning that there is an operator $B'$ acting on ${\cal H}$ having the property that 
\begin{equation}
\label{m43}
\Psi_0 = \Psi_\s\otimes\Omega_\r = B' \Omega_{\s\r,\beta,\lambda} \quad \mbox{and $B'$ commutes with  $e^{\i tL} (X\otimes\bbbone_\s\otimes\bbbone_\r)e^{-\i tL}$}
\end{equation}
for all system observables $X$. Moreover,
\begin{equation}
\label{m44}
B' = \bbbone_\s\otimes B\otimes\bbbone_\r +O(\lambda^2)
\end{equation}
(see e.g. \cite{KM-CP}, Lemma 3.4). Concretely, the operator $B$ is obtained solving the relation
\begin{equation}
\label{m45}
\Psi_\s = \big(\bbbone_\s\otimes B\big) \Omega_{\s,\beta},
\end{equation}
which has a unique solution for any given $\Psi_\s$. 

\medskip
{\em Example.} Take the pure initial system state in which each donor level is populated equally, with probability $1/N_D$. The corresponding purified vector state is $\Psi_\s =|D\rangle\otimes |D\rangle$, see \eqref{Dstate'}. We show how to solve \eqref{m45} for $B$. We will find a $B$ acting nontrivially only on $\bar{\cal H}_\s$, i.e., satisfying $B\xi_{\alpha,j}=0$ (see \eqref{m38}). We expand $|D\rangle$ in the eigenbasis $\{\varphi_1,\varphi_2\}$ (see \eqref{m22}), $|D\rangle = x_1|\varphi_1\rangle +x_2|\varphi_2\rangle$, where $x_1$, $x_2\in\mathbb R$. Thus we need to solve the following equation for $B$,
\begin{equation}
x^2_1 \varphi_{11} + x_1x_2\big(\varphi_{12} +\varphi_{21} \big) +x_2^2 \varphi_{21} =  \big(\bbbone_\s\otimes B\big) (\alpha_1 \varphi_{11} +\alpha_2\varphi_{22} ),
\label{m65}
\end{equation}
where $\varphi_{ij}=\varphi_i\otimes\varphi_j$ and $\alpha_j = Z_{\s,\beta}^{-1/2} e^{-\beta e_j/2}$. Taking $\langle \varphi_1|\otimes\bbbone_\s$ on both sides of \eqref{m65} gives $x^2_1\varphi_1+x_1x_2\varphi_2 = \alpha_1 B\varphi_1$ and similarly we get $x_1x_2\varphi_1 +x^2_2 \varphi_1 = \alpha_2 B\varphi_2$. Hence, as a matrix in the basis $\{\varphi_1,\varphi_2\}$, we get
\begin{eqnarray}
B &=& 
\begin{pmatrix}
	x^2_1/\alpha_1 & x_1x_2/\alpha_1\\
	x_1x_2/\alpha_2 & x^2_2/\alpha_2
\end{pmatrix}
= Z_{\s,\beta}^{1/2} 
\begin{pmatrix}
	x^2_1 e^{\beta e_1/2}  & x_1x_2 e^{\beta e_1/2}\\
	x_1x_2 e^{\beta e_2/2}& x^2_2e^{\beta e_2/2}
\end{pmatrix}\nonumber\\
&=&  Z_{\s,\beta}^{1/2} \, e^{\beta \bar H_\s/2}
\begin{pmatrix}
x^2_1  & x_1x_2 \\
x_1x_2& x^2_2
\end{pmatrix} = Z_{\s,\beta}^{1/2} \, e^{\beta \bar H_\s/2} \big| x_1\varphi_1 +x_2\varphi_2\big\rangle \big\langle x_1 \varphi_1 +x_2\varphi_2\big| \nonumber\\
&=& Z_{\s,\beta}^{1/2} \, e^{\beta \bar H_\s/2} |D\rangle\langle D|.  
\label{m66}
\end{eqnarray}
Even though this $B$ is determined explicitly in \eqref{m66} we will see below that we can find the dynamics without using the specific form of $B$ (see for instance \eqref{ce}). 
\medskip

We combine \eqref{m43}-\eqref{m45}  with \eqref{m19} to obtain
\begin{eqnarray}
\av{X}_t &=& \scalprod{\Psi_0}{e^{\i tL} (X\otimes\bbbone_\s\otimes\bbbone_\r)e^{-\i tL}\Psi_0}\nonumber\\
&=&  \scalprod{\Psi_0}{ B' e^{\i tL} (X\otimes\bbbone_\s\otimes\bbbone_\r)\Omega_{\s\r,\beta,\lambda}}\nonumber\\
&=& \scalprod{\Psi_0}{ (\bbbone_\s\otimes B\otimes\bbbone_\r)  e^{\i tL} (X\otimes\bbbone_\s\otimes\bbbone_\r)\ \Omega_{\s,\beta}\otimes\Omega_\r}  +O(\lambda^2),
\label{m41}
\end{eqnarray}
where the remainder is uniform in $t$. We have used in the second step that $L\Omega_{\s\r,\beta,\lambda}=0$ and in the last one that 
\begin{equation}
\label{m39}
\Omega_{\s\r,\beta,\lambda} = \Omega_{\s,\beta}\otimes\Omega_\r + O(\lambda^2).
\end{equation}
Writing for short 
\begin{equation}
\label{forshort}
B\equiv \bbbone_\s\otimes B\otimes\bbbone_\r\mbox{\ \ and\ \ } X\equiv X\otimes\bbbone_\s\otimes\bbbone_\r,
\end{equation} 
we obtain from \eqref{m41} that 
\begin{eqnarray}
\av{X}_t &=& \scalprod{\Psi_\s\otimes\Omega_\r}{\big( B e^{\i tL} X\big)\Omega_{\s,\beta}\otimes\Omega_\r} +O(\lambda^2)\nonumber\\
&=&  \sum_{j=1}^4 \scalprod{\Psi_\s\otimes\Omega_\r}{\big( B e^{\i tL_j} P_j X\big)\Omega_{\s,\beta}\otimes\Omega_\r} +O(\lambda^2),
\label{m42}
\end{eqnarray}
where we have used the reduction \eqref{m10.1} of the dynamics and 
where $P_j$ is the orthogonal projection onto ${\cal H}_j$, c.f. \eqref{m10}. The {\em dynamical resonance representation} now gives a concrete perturbation expansion for each propagator $e^{\i tL_j}$ (c.f. \cite{MSB, MSBPRL, Mdyn, KM-CP, KM-dyn}). Namely, in \eqref{m42} we can use the expansion
\begin{equation}
\label{m42.1}
e^{\i t L_j} =  \sum_{s=1}^{4} e^{\i t \varepsilon_j^{(s)}} \Pi_j^{(s)} +O(\lambda^2),
\end{equation}
where the remainder is bounded independently of time $t$. 
Here, $j$ labels the invariant sectors and $s$ ranges over the number of distinct eigenvalues and resonances $\varepsilon_j^{(s)}$ of $L_j$ (there happen to be four of them for each $j$, see Lemmas \ref{lemma2}-\ref{lem1}). Here $\Pi_j^{(s)}$ is the  eigenprojection (resonance projection) associated to $\varepsilon_j^{(s)}$, given explicitly in Lemmas \ref{lemma2}-\ref{lemma4} and which we now list {\em modulo $O(\lambda^2)$ terms}. An equality leaving away the $O(\lambda^2)$ errors, which are uniform in $t\ge 0$, is indicated by $\doteq$\, . From Lemma \ref{lemma2},
\begin{eqnarray}
\Pi_1^{(1)} &\doteq& |\bar\Omega_{\s,\beta}\rangle\langle\bar\Omega_{\s,\beta}|\otimes|\Omega_\r\rangle\langle \Omega_\r|,  \qquad \mbox{c.f. \eqref{m46}}\nonumber\\
\Pi_1^{(2)} &\doteq&  |\Psi_1^{(2)}\rangle\langle \Psi_1^{(2)} |\otimes|\Omega_\r\rangle\langle \Omega_\r|, \mbox{\qquad c.f. \eqref{m47}}\nonumber\\
\Pi_1^{(3)} &\doteq& |\varphi_{12}\rangle\langle\varphi_{12}|\otimes|\Omega_\r\rangle\langle \Omega_\r|,\nonumber\\
\Pi_1^{(4)} &\doteq& |\varphi_{21}\rangle\langle\varphi_{21}|\otimes|\Omega_\r\rangle\langle \Omega_\r|.
\label{m49}
\end{eqnarray}
Lemma \ref{lemma3} gives  ({\em c.f.} \eqref{m30})
\begin{equation}
\Pi_2^{(s)} \doteq 
\left\{
\begin{array}{ll}
|\varphi_s\rangle\langle\varphi_s|\otimes P_{D\perp}\otimes|\Omega_\r\rangle\langle\Omega_\r|, & s=1,2, \nonumber\\
|\varphi_{s-2}\rangle\langle\varphi_{s-2}|\otimes P_{A\perp}\otimes|\Omega_\r\rangle\langle\Omega_\r|, &s=3,4 . 
\end{array}
\right.
\label{m50}
\end{equation}
From Lemma \ref{lemma4} we get ({\em c.f.} \eqref{m55})
\begin{equation}
\Pi_3^{(s)} \doteq 
\left\{
\begin{array}{ll}
P_{D\perp}\otimes|\varphi_s\rangle\langle\varphi_s|\otimes |\Omega_\r\rangle\langle\Omega_\r|, & s=1,2, \\
P_{A\perp}\otimes |\varphi_{s-2}\rangle\langle\varphi_{s-2}|\otimes |\Omega_\r\rangle\langle\Omega_\r|, & s=3,4 .
\end{array}
\right.
\label{m58}
\end{equation}
Finally, Lemma \ref{lem1} shows that, for arbitrary $\lambda\in{\mathbb R}$  ({\em c.f.}  \eqref{m59})
\begin{eqnarray}
\Pi_4^{(1)} &=& P_{D\perp}\otimes P_{D\perp}\otimes |\Psi_D\rangle\langle \Psi_D|,  \nonumber\\
\Pi_4^{(2)} &=& P_{A\perp}\otimes P_{A\perp}\otimes |\Psi_A\rangle\langle \Psi_A|, \nonumber\\
\Pi_4^{(3)} &=& P_{D\perp}\otimes P_{A\perp}\otimes |\Psi_{DA}\rangle\langle \Psi_{DA}|, \nonumber\\
\Pi_4^{(4)} &=& P_{A\perp}\otimes P_{D\perp}\otimes |\Psi_{AD}\rangle\langle \Psi_{AD}|.
\label{m60}
\end{eqnarray}
The expressions for $\Pi_4^{(j)}$, $j=1,\ldots,4$, are exact equalities, there are no remainders in the four relations \eqref{m60}.

We combine \eqref{m42} and \eqref{m42.1} and use the expressions \eqref{m49}-\eqref{m60}. We view $\Pi_j^{(s)}$, for $j=1,2,3$, as operators on ${\cal H}_\s\otimes{\cal H}_\s$ ({\em i.e.}, we leave out the factor  $|\Omega_\r\rangle\langle\Omega_\r|$, {\em c.f.} \eqref{m49}-\eqref{m58}). For $j=4$ we introduce the notation $\widetilde \Pi_4^{(j)}$ to denote the `part on the DA space', for instance (c.f. \eqref{m60})
\begin{equation}
\widetilde \Pi_4^{(1)} = P_{D\perp}\otimes P_{D\perp}.
\label{before}
\end{equation} 
We arrive at the following result. 
\begin{prop}
\label{prop3.1}
Let $\Psi_\s$ and $X\in {\mathcal B}({\mathcal H}_\s)$ be any initial DA state and any DA operator (observable). Then
\begin{equation}
\av{X}_t = \sum_{j=1}^3\sum_{s=1}^4 e^{\i t\varepsilon_j^{(s)}}   \scalprod{\Psi_\s}{ B \Pi_j^{(s)} X \Omega_{\s,\beta}}
+\sum_{s=1}^4  e^{\i t\varepsilon_4^{(s)}} \scalprod{\Psi_\s}{ B \widetilde\Pi_4^{(s)} X \Omega_{\s,\beta}} 
 + O(\lambda^2),
 \label{m61}
 \end{equation}
with a remainder independent of time $t\ge 0$. 
\end{prop}

{\em Remark. } According to \eqref{m60}, the last sum in \eqref{m61} is actually
\begin{equation}
	\label{m117}
\sum_{s=1}^4 w_s \, e^{\i t\varepsilon_4^{(s)}} \scalprod{\Psi_\s}{ B \widetilde\Pi_4^{(s)} X \Omega_{\s,\beta}}, 
\end{equation}
with weights 
$w_1 = |\scalprod{\Psi_D}{\Omega_\r}|^2$, $w_2 = |\scalprod{\Psi_A}{\Omega_\r}|^2$, $w_3 = |\scalprod{\Psi_{DA}}{\Omega_\r}|^2$, $w_4 = |\scalprod{\Psi_{AD}}{\Omega_\r}|^2$, 
where the  $\Psi_X$ are defined in \eqref{m14} and \eqref{psida}. The $w_j$ depend on $\lambda$ and satisfy $w_j=1+O(\lambda^2)$. Since the remainder in \eqref{m61} is already $O(\lambda^2)$ we can replace $w_j$ by $1$. One can also calculate  $w_j$ exactly, as we illustrate now. 

{\em Explicit form of $w_1$}:  Call for short $f=\i\lambda E_Dh/\omega$, so $w_1 = |\scalprod{W(f)J_\r W(f)J_\r\Omega}{\Omega}|^2$. Remembering that we are in the thermal representation of the quantum field, the Weyl operators $W(f)$ are given by $e^{\i \varphi(f_\beta)}$, where test function $f_\beta$ is defined by \eqref{aw}. The effect of conjugating with $J_\r$ is (\cite{BR, MSB}):
\begin{equation}
	\label{m114}
J_\r W(f_\beta(u,\Sigma))J_\r = W(\overline f_\beta(-u,\Sigma)) = W(-e^{-\beta u/2} f_\beta(u,\Sigma)).
\end{equation}
Then, using the CCR, $W(f)W(g)= e^{-\frac \i2{\rm Im} \langle f,g\rangle} W(f+g)$, together with $\scalprod{\Omega_\r}{W(h)\Omega_\r} = e^{-\frac14 \| h\|^2}$ (for $f,g,h \in L^2({\mathbb R}\times S^2, du\times d\Sigma)$), one readily obtains
\begin{eqnarray}
w_1 &=& \big|\scalprod{ \Omega_\r} { W\big( (1-e^{-\beta u/2} ) f_\beta \big) \Omega_\r }\big|^2\nonumber\\
&=&\exp -\tfrac12\int_{{\mathbb R}\times S^2}  \frac{(1-e^{-\beta u/2})^2}{|1-e^{-\beta u}|} |f(|u|,\Sigma)|^2 u^2 du d\Sigma\nonumber\\
&=& \exp -\int_{ {\mathbb R}^3 }  \tanh(\beta |k|/4) |f(k)|^2  d^3k\nonumber\\
&=&\exp -\lambda^2 E_D^2 \int_{{\mathbb R^3}}  \frac{\tanh(\beta |k|/4)}{|k|} |h(k)|^2 d^3k.
\label{w1}
\end{eqnarray}
In \eqref{w1}, the function $h$ is the form factor of the interaction, \eqref{hamiltonian}. 
Similarly one obtains $w_2$ and one can also calculate $w_{3,4}$ explicitly along these lines.

\subsection{Proofs of Theorem \ref{dynthm} and Proposition \ref{corprop}}

\bigskip

{\bf Proof of Theorem \ref{dynthm}. } To derive the result we use the expansion \eqref{m61} given in Proposition \ref{prop3.1} and analyze the individual terms in this expansion.  
\bigskip

$\circ$ Consider $j=1$ and $s=1$. We have 
\begin{equation}
\scalprod{\Psi_\s}{ B \Pi_1^{(1)} X \Omega_{\s,\beta}} = \scalprod{\Psi_\s}{ (\bbbone_\s\otimes B)\bar\Omega_{\s,\beta}} \scalprod{\bar\Omega_{\s,\beta}}{ ( X\otimes\bbbone_\s) \Omega_{\s,\beta}}
\label{m68}
\end{equation}
and using the expressions of $\bar\Omega_{\s,\beta}$ and $\Omega_{\s,\beta}$ given in \eqref{m46} and \eqref{m38}, we obtain
\begin{eqnarray}
\scalprod{\bar\Omega_{\s,\beta}}{ ( X\otimes\bbbone_\s) \Omega_{\s,\beta}} &=& Z^{-1/2}_{\s,\beta} \sum_{j=1,2} e^{-\beta e_j/2} \scalprod{\bar\Omega_{\s,\beta}}{ ( X\otimes\bbbone_\s) \varphi_j\otimes\varphi_j} \nonumber\\
& =& c^{-1}\, {\rm Tr}\big( \bar\rho_{\s,\beta} \bar P_\s X \bar P_\s\big),
\label{m99}
\end{eqnarray}
where we have taken into account that $\varphi_j\perp \xi_{\alpha,\ell}$ in \eqref{m38} and we have introduced 
\begin{equation}
c =\sqrt{\frac{{\rm Tr} e^{-\beta H_\s}}{{\rm Tr} e^{-\beta \bar H_\s}}}.
\label{ce}
\end{equation}
Next, it follows from \eqref{m38} that 
\begin{equation}
\varphi_{k\ell} \equiv \varphi_k\otimes\varphi_\ell = \sqrt{{\rm Tr} e^{-\beta H_\s}}\ e^{\beta e_\ell/2}\ \big( |\varphi_k\rangle\langle\varphi_\ell | \otimes \bbbone_\s\big) \Omega_{\s,\beta},\qquad k,\ell\in \{1,2\}.
\label{m1}
\end{equation}
Due to \eqref{m45}, this implies the relation
\begin{equation}
(\bbbone_\s\otimes B) \varphi_{k\ell} = \sqrt{{\rm Tr} e^{-\beta H_\s}}\ e^{\beta e_\ell/2}\ \big( |\varphi_k\rangle\langle\varphi_\ell | \otimes \bbbone_\s\big) \Psi_\s, \qquad k,\ell\in \{1,2\}.
\label{m2}
\end{equation}
Therefore, we obtain from \eqref{m46}  
\begin{eqnarray}
\lefteqn{\scalprod{\Psi_\s}{ (\bbbone_\s\otimes B)\bar\Omega_{\s,\beta}}}\nonumber\\
&=& \frac{e^{-\beta e_1/2}}{\sqrt{e^{-\beta e_1} +e^{-\beta e_2}}}\scalprod{\Psi_\s}{ (\bbbone_\s\otimes B)\varphi_{11}} +\frac{e^{-\beta e_2/2}}{\sqrt{e^{-\beta e_1} +e^{-\beta e_2}}}\scalprod{\Psi_\s}{ (\bbbone_\s\otimes B)\varphi_{22}}\nonumber\\
&=& c\  \big( [\rho_0]_{11} + [\rho_0]_{22}\big),
\end{eqnarray}
where 
\begin{equation}
[\rho_0]_{k\ell} = \scalprod{\varphi_k}{\rho_0\varphi_\ell} = \scalprod{\Psi_\s}{(|\varphi_\ell\rangle\langle\varphi_k|\otimes\bbbone_\s) \Psi_\s}.
\label{m130}
\end{equation}

Combining \eqref{m68} with \eqref{m99} and \eqref{m130} gives 
\begin{equation}
\label{m110}
\scalprod{\Psi_\s}{ B \Pi_1^{(1)} X \Omega_{\s,\beta}} = \big( [\rho_0]_{11} + [\rho_0]_{22}\big)\ 
{\rm Tr}\big( \bar\rho_{\s,\beta} \bar P_\s X \bar P_\s\big).
\end{equation}

$\circ$ Next we look at $j=1$, $s=2$. Using the definition \eqref{m47} gives
\begin{eqnarray}
\lefteqn{
	\scalprod{\Psi_\s}{(\bbbone_\s\otimes B)\Psi^{(2)}_1} }\nonumber\\
&=& \frac{e^{\beta e_1/2}}{\sqrt{e^{\beta e_1} +e^{\beta e_2}}}\scalprod{\Psi_\s}{ (\bbbone_\s\otimes B)\varphi_{11}} -\frac{e^{\beta e_2/2}}{\sqrt{e^{\beta e_1} +e^{\beta e_2}}}\scalprod{\Psi_\s}{ (\bbbone_\s\otimes B)\varphi_{22}}\nonumber\\
&=& \sqrt{ \frac{ {\rm Tr} e^{-\beta H_\s} }{e^{\beta e_1} +e^{\beta e_2}}} \big( e^{\beta e_1} [\rho_0]_{11} - e^{\beta e_2}[\rho_0]_{22}\big)\nonumber\\
&=& c e^{-\beta( e_1+e_2)/2} \big( e^{\beta e_1} [\rho_0]_{11} - e^{\beta e_2}[\rho_0]_{22}\big).
\label{m80g}
\end{eqnarray}
Combining \eqref{m80g} with
$$
\scalprod{\Psi^{(2)}_1}{(X\otimes\bbbone_\s)\Omega_{\s,\beta}} = c^{-1} \scalprod{\Psi^{(2)}_1}{(X\otimes\bbbone_\s)\bar\Omega_{\s,\beta}} = \frac{ \scalprod{\varphi_1}{X\varphi_1} -  \scalprod{\varphi_2}{X\varphi_2}}{c\,  \sqrt{(e^{\beta e_1} +e^{\beta e_2}) (e^{-\beta e_1} +e^{-\beta e_2})}}
$$
yields (use $\sqrt{ \frac{e^{-\beta e_1} +e^{-\beta e_2}}{e^{\beta e_1}+e^{\beta e_2}}} = e^{-\beta(e_1+e_2)/2}$)
\begin{eqnarray}
\scalprod{\Psi_\s}{(\bbbone_\s\otimes B) \Pi^{(2)}_1 X\Omega_{\s,\beta}} &=& \frac{ e^{-\beta e_2} [\rho_0]_{11} - e^{-\beta e_1}[\rho_0]_{22}}{e^{-\beta e_1} +e^{-\beta e_2}}\big( \scalprod{\varphi_1}{X\varphi_1} -  \scalprod{\varphi_2}{X\varphi_2}\big).\qquad \quad
\label{m81g}
\end{eqnarray}

$\circ$ Next we address the cases $j=1$, $s=3, 4$. Using \eqref{m2} we obtain
\begin{equation}
\scalprod{\Psi_\s}{(\bbbone_\s\otimes B)\varphi_{12}} = e^{\beta e_2/2} \sqrt{{\rm Tr} e^{-\beta  H_\s}}\ [\rho_0]_{21}.
\label{m83g}
\end{equation}
Combining \eqref{m83g} with
\begin{equation}
\scalprod{\varphi_{12}}{(X\otimes\bbbone_\s) \Omega_{\s,\beta}} = \frac{e^{-\beta e_2/2} \scalprod{\varphi_1}{X\varphi_2}}{\sqrt{{\rm Tr}e^{-\beta H_\s}} }
\label{m82}
\end{equation}
gives 
\begin{equation}
\scalprod{\Psi_\s}{(\bbbone_\s\otimes B) \Pi_1^{(3)} X\Omega_{\s,\beta}} = [\rho_0]_{21} \scalprod{\varphi_1}{X\varphi_2}.
\label{m84g}
\end{equation}
The case $s=4$ is addressed just like $s=3$, with the result
\begin{equation}
\scalprod{\Psi_\s}{(\bbbone_\s\otimes B) \Pi_1^{(4)} X\Omega_{\s,\beta}} = [\rho_0]_{12} \scalprod{\varphi_2}{X\varphi_1}.
\label{m85g}
\end{equation}

$\circ$ Now we consider $j=2$, $s=1,\ldots,4$. We need to analyze
\begin{eqnarray}
\scalprod{\Psi_\s}{(\bbbone_\s\otimes B) \Pi_2^{(1)} (X\otimes\bbbone_\s) \Omega_{\s,\beta}} &=&\scalprod{\Psi_\s}{(\bbbone_\s\otimes B) (|\varphi_1\rangle\langle\varphi_1|\otimes P_{D\perp} )(X\otimes \bbbone_\s) \Omega_{\s,\beta}} \nonumber\\
&=& \scalprod{\Psi_\s}{ \{(|\varphi_1\rangle\langle\varphi_1| X)\otimes( B P_{D\perp} )  \}\Omega_{\s,\beta}}\nonumber\\
&=& \scalprod{\Psi_\s}{ \{(|\varphi_1\rangle\langle\varphi_1| XP_{D\perp})\otimes B    \}\Omega_{\s,\beta}}\nonumber\\
&=& \scalprod{\Psi_\s}{ (|\varphi_1\rangle\langle\varphi_1| XP_{D\perp})\Psi_\s}\nonumber\\
&=& {\rm Tr} \big( \rho_0|\varphi_1\rangle\langle\varphi_1| XP_{D\perp}\big).\qquad \quad
\label{m102g}
\end{eqnarray}
In the third step we have used that $(\bbbone_\s\otimes P_{D\perp})\Omega_{\s,\beta} = (P_{D\perp}\otimes\bbbone_\s )\Omega_{\s,\beta}$ (see \eqref{m38}) and in the fourth one that $(\bbbone_\s\otimes B)\Omega_{\s,\beta}=\Psi_\s$.

Proceeding in the same way for $\Pi_2^{(2)}$ we readily obtain
\begin{equation}
\scalprod{\Psi_\s}{(\bbbone_\s\otimes B)\Pi_2^{(2)}(X\otimes \bbbone_\s)\Omega_{\s,\beta}} ={\rm Tr} \big( \rho_0|\varphi_2\rangle\langle\varphi_2| XP_{D\perp}\big).
\label{m104g}
\end{equation}
Finally, for $\Pi_2^{(s)}$ with $s=3,4$, the same analysis holds and we get
\begin{equation}
\scalprod{\Psi_\s}{(\bbbone_\s\otimes B)\Pi_2^{(s)}(X\otimes \bbbone_\s)\Omega_{\s,\beta}} =
\left\{
\begin{array}{ll}
{\rm Tr} \big( \rho_0|\varphi_1\rangle\langle\varphi_1| XP_{A\perp}\big), & s=3,\\
{\rm Tr} \big( \rho_0|\varphi_2\rangle\langle\varphi_2| XP_{A\perp}\big), & s=4.
\end{array}
\right.
\label{m105g}
\end{equation}

$\circ$ Consider $j=3$, $s=1,\dots,4$. Proceeding as for $j=2$ and also using that $(\bbbone_\s\otimes |\varphi_\ell\rangle\langle \varphi_\ell|)\Omega_{\s,\beta} = ( |\varphi_\ell\rangle\langle \varphi_\ell|\otimes \bbbone_\s)\Omega_{\s,\beta}$, $\ell=1,2$ (see \eqref{m38}) we readily get
\begin{equation}
\scalprod{\Psi_\s}{(\bbbone_\s\otimes B) \Pi_3^{(s)}(X\otimes \bbbone_\s)\Omega_{\s,\beta}} =
\left\{ 
\begin{array}{ll}
{\rm Tr}\big(\rho_0 P_{D\perp}X|\varphi_s\rangle\langle\varphi_s|\big), & s=1,2,\\
{\rm Tr}\big(\rho_0 P_{A\perp}X|\varphi_{s-2}\rangle\langle\varphi_{s-2}|\big), & s=3,4.
\end{array}
\right.
\label{m106g}
\end{equation}

$\circ$ Take now $j=4$. Proceeding in a by now standard way, as above, we get
\begin{eqnarray}
\scalprod{\Psi_\s}{(\bbbone_\s\otimes B)\widetilde\Pi_4^{(s)} (X\otimes \bbbone_\s)\Omega_{\s,\beta}} =
\left\{ 
\begin{array}{ll}
{\rm Tr}\big( \rho_0 P_{D\perp} X P_{D\perp}\big) & s=1\\
{\rm Tr}\big( \rho_0 P_{A\perp} X P_{A\perp}\big) & s=2\\
{\rm Tr}\big( \rho_0 P_{D\perp} X P_{A\perp}\big) & s=3\\
{\rm Tr}\big( \rho_0 P_{A\perp} X P_{D\perp}\big) & s=4\\
\end{array}
\right.
.
\label{m108g}
\end{eqnarray}

The result of Proposition \ref{prop3.1} together with the relations \eqref{m110}, \eqref{m81g}, \eqref{m84g}, \eqref{m85g}, \eqref{m102g}, \eqref{m104g}, \eqref{m105g}, \eqref{m106g}, \eqref{m108g}  implies
\begin{eqnarray}
	\langle X\rangle_t &=& \big( [\rho_0]_{11} + [\rho_0]_{22}\big)\ 
	{\rm Tr}\big( \bar\rho_{\s,\beta} \bar P_\s X \bar P_\s\big)	\label{dynamics'}\\
	&&+ {\rm Tr}\big( \rho_0 P_{D\perp} X P_{D\perp}\big) + {\rm Tr}\big( \rho_0 P_{A\perp} X P_{A\perp}\big)\nonumber \\
	&& + e^{i t \varepsilon_4^{(3)}} {\rm Tr}\big( \rho_0 P_{D\perp} X P_{A\perp}\big)+ e^{i t\varepsilon_4^{(4)}}
	{\rm Tr}\big( \rho_0 P_{A\perp} X P_{D\perp}\big)\nonumber\\
	&& + e^{it \varepsilon_1^{(2)}} \ \frac{ e^{-\beta e_2} [\rho_0]_{11} - e^{-\beta e_1}[\rho_0]_{22}}{e^{-\beta e_1} +e^{-\beta e_2}}\big( \scalprod{\varphi_1}{X\varphi_1} -  \scalprod{\varphi_2}{X\varphi_2}\big)\nonumber\\
	&& + e^{it \varepsilon_1^{(3)}} \ [\rho_0]_{21} \scalprod{\varphi_1}{X\varphi_2} + e^{-it (\varepsilon_1^{(3)})^* } \ [\rho_0]_{12} \scalprod{\varphi_2}{X\varphi_1}\nonumber\\
	&& + \sum_{s=1,2} e^{it \varepsilon_2^{(s)}} {\rm Tr} \big( \rho_0|\varphi_s\rangle\langle\varphi_s| XP_{D\perp}\big)+\sum_{s=3,4} e^{it \varepsilon_2^{(s)}}   {\rm Tr} \big( \rho_0|\varphi_{s-2} \rangle\langle\varphi_{s-2} | XP_{A\perp}\big)\nonumber\\
	&& +\sum_{s=1,2}e^{it \varepsilon_3^{(s)}}  {\rm Tr}\big(\rho_0 P_{D\perp}X|\varphi_s\rangle\langle\varphi_s|\big)
	+\sum_{s=3,4}e^{it \varepsilon_3^{(s)}}
	{\rm Tr}\big(\rho_0 P_{A\perp}X|\varphi_{s-2}\rangle\langle\varphi_{s-2}|\big)\nonumber\\
	&& +O(\lambda^2),\nonumber
\end{eqnarray}
where the remainder term is uniform in $t\ge 0$. Now since $\{\varphi_1,\varphi_2\}$ is an orthonormal basis of $\bar{\mathcal H}_\s = {\rm Ran}\bar P_\s$, we have $[\rho_0]_{11}+[\rho_0]_{22}= {\rm Tr}(\bar P_\s \rho_0 )$, so the first term on the right side of \eqref{dynamics} is ${\rm Tr}(\rho_0\bar P_\s) {\rm Tr}( \bar P_\s \bar \rho_{\s,\beta} \bar P_\s X)$, which gives the first contribution to the right side of \eqref{dynamics}. The other ones are obtained similarly from \eqref{dynamics'}. This  concludes the proof of  Theorem \ref{dynthm}. \hfill \qed

\bigskip

{\bf Proof of Proposition \ref{corprop}. } This is a direct calculation of the right hand side of \eqref{dynamics''}, taking into account the relations
\begin{eqnarray}
\langle D_k| P_{11}\rho_0 P_{11} | D_\ell\rangle &=& \frac{1}{N_D}\frac{1}{1+\alpha^2} [\rho_0]_{11},\nonumber\\
\langle D_k| P_{12}\rho_0 P_{21} | D_\ell\rangle &=& \frac{1}{N_D}\frac{1}{1+\alpha^2} [\rho_0]_{22},\nonumber\\
\langle D_k| P_{22}\rho_0 P_{22} | D_\ell\rangle &=& \frac{1}{N_D}\frac{\alpha^2}{1+\alpha^2} [\rho_0]_{22},\nonumber\\
\langle D_k| P_{21}\rho_0 P_{12} | D_\ell\rangle &=& \frac{1}{N_D}\frac{\alpha^2}{1+\alpha^2} [\rho_0]_{11},\nonumber\\
\langle D_k| P_{D\perp}\rho_0 P_{D\perp} | D_\ell\rangle &=& \langle D_k|\rho_0 |D_\ell\rangle +\frac{1}{N_D} \langle D|\rho_0|D\rangle\nonumber\\
&& -\frac{1}{\sqrt N_D}  \big(\langle D_k|\rho_0|D\rangle +\langle D|\rho_0| D_\ell\rangle \big),\nonumber\\
P_{A\perp} |D_k\rangle &=& P_{A\perp} |D_\ell\rangle=0,\nonumber\\
\langle D_k| P_{11}\rho_0 P_{22} | D_\ell\rangle &=& \frac{1}{N_D}\frac{|\alpha|}{1+\alpha^2} [\rho_0]_{21},\nonumber\\
\langle D_k| P_{D\perp}\rho_0 P_{ss} | D_\ell\rangle &=& 0.
\label{m151}
\end{eqnarray}
Remark that we can also compute the matrix elements between acceptor levels, and those between donor and acceptor levels.\hfill \qed

\section{Conclusion}

We consider a donor-acceptor (DA) system described by $N_D$ and $N_A$ sites at the degenerate energies $E_D$ and $E_A$. Each donor site is coupled equally to each acceptor site, and both the donor and acceptor are coupled to a common noise, modeled by a thermal Bose field of vibrations. We use the dynamical resonance theory  to find the effective evolution of the DA system for all times $t\ge 0$, up to an error term which vanishes quadratically in the DA-reservoir coupling (independently of time). Due to the symmetry of the Hamiltonian, the dynamics has many stationary states. We exhibit them explicitly. We show that the DA final state (time $t\rightarrow\infty$) depends on its initial state ($t=0$) and we find the initial-final state correspondence.  The amount of population transferred from the donor to the acceptor during the process depends on quantum properties of the initial donor state: we demonstrate that if the initial population is shared coherently by the donor sites then the transfer to the acceptor is high. For an incoherently populated donor, the transfer is low.  We examine the fluctuations in the donor populations during the transfer process and show that they decrease with increasing system size $N_D$. We also discuss, in a qualitative way, what will change in our results when the symmetry (degeneracy) of the Hamiltonian is lifted.

\bigskip

{\bf Acknowledgement.\ }
M.M. was supported by a Discovery Grant from the National Sciences and Engineering Research Council of Canada (NSERC)  and in part by funding from the Simons
Foundation and the Centre de Recherches Math\'ematiques, through the Simons-CRM
scholar-in-residence program. 
The work by G.P.B. and A.S. was done at Los Alamos National Laboratory managed by Triad National Security,
LLC, for the National Nuclear Security Administration of the U.S. Department of Energy under Contract No.
89233218CNA000001.

\end{document}